\documentclass[a4paper,fleqn]{cas-dc}
\usepackage[authoryear]{natbib}
\usepackage[figuresright]{rotating}
\usepackage{moreverb}
\usepackage{amsmath,amssymb,amsfonts}
\usepackage{algorithmic}
\usepackage{url}
\usepackage{graphicx}
\usepackage{textcomp}
\usepackage{xcolor}
\usepackage{ragged2e}
\usepackage{balance}
\usepackage{colortbl}
\usepackage{multirow}
\usepackage{color}

\usepackage{CJKutf8}
\usepackage{siunitx}
\usepackage{rotfloat}
\usepackage{float}
\usepackage{lscape}
\usepackage{amssymb,mathrsfs,amsmath}

\def\tsc#1{\csdef{#1}{\textsc{\lowercase{#1}}\xspace}}
\tsc{WGM}
\tsc{QE}
\tsc{EP}
\tsc{PMS}
\tsc{BEC}
\tsc{DE}

\begin{document}
\begin{sloppypar}

\let\WriteBookmarks\relax
\def\floatpagepagefraction{1}
\def\textpagefraction{.001}
\shorttitle{Bug Priority Change}
\shortauthors{Z Li et al.}
\title [mode = title]{Bug Priority Change: An Empirical Study on Apache Projects}

\author[1]{Zengyang Li}
\ead{zengyangli@ccnu.edu.cn}

\credit{Conceptualization of this study, Methodology, Investigation, Data curation, Writing - Original draft preparation}

\address[1]{School of Computer Science \& Hubei Provincial Key Laboratory of Artificial Intelligence and Smart Learning, \\Central China Normal University, Wuhan, China \\}

\author[1]{Guangzong Cai}
\ead{guangzongcai@mails.ccnu.edu.cn}
\credit{Investigation, Data curation, Software, Writing - Original draft preparation}

\author[1]{Qinyi Yu}
\ead{qinyiyu@mails.ccnu.edu.cn}
\credit{Investigation, Data curation, Software, Writing - Original draft preparation}

\author[2]{Peng Liang}
\cormark[1]
\ead{liangp@whu.edu.cn}
\credit{Conceptualization of this study, Methodology, Writing - Original draft preparation}
\address[2]{School of Computer Science, Wuhan University, Wuhan, China}

\author[1]{Ran Mo}
\ead{moran@ccnu.edu.cn}
\credit{Methodology, Writing - Original draft preparation}

\author[3]{Hui Liu}
\ead{hliu@hust.edu.cn}
\credit{Methodology, Writing - Original draft preparation}
\address[3]{School of Artificial Intelligence and Automation, Huazhong University of Science and Technology, Wuhan, China}

\cortext[cor1]{Corresponding author.}


\begin{abstract}
In issue tracking systems, each bug is assigned a priority level (e.g., Blocker, Critical, Major, Minor, or Trivial in JIRA from highest to lowest), which indicates the urgency level of the bug. In this sense, understanding bug priority changes helps to arrange the work schedule of participants reasonably, and facilitates a better analysis and resolution of bugs. According to the data extracted from JIRA deployed by Apache, a proportion of bugs in each project underwent priority changes after such bugs were reported, which brings uncertainty to the bug fixing process. However, there is a lack of in-depth investigation on the phenomenon of bug priority changes, which may negatively impact the bug fixing process. Thus, we conducted a quantitative empirical study on bugs with priority changes through analyzing 32 non-trivial Apache open source software projects. The results show that: (1) 8.3\% of the bugs in the selected projects underwent priority changes; (2) the median priority change time interval is merely a few days for most (28 out of 32) projects, and half (50. 7\%) of bug priority changes occurred before bugs were handled; (3) for all selected projects, 87.9\% of the bugs with priority changes underwent only one priority change, most priority changes tend to shift the priority to its adjacent priority, and a higher priority has a greater probability to undergo priority change; (4) bugs that require bug-fixing changes of higher complexity or that have more comments are likely to undergo priority changes; and (5) priorities of bugs reported or allocated by a few specific participants are more likely to be modified, and maximally only one participant in each project tends to modify priorities.
\end{abstract}


\maketitle

\section{Introduction}
\label{chap:intro}
Bug fixing is an important maintenance activity in the software development process. In issue tracking systems, each bug is assigned a priority (e.g., Blocker, Critical, Major, Minor, or Trivial in JIRA from the highest to lowest), which indicates the urgency level of the bug \citep{TiAlLo2016}. In this sense, understanding bug priority changes helps to arrange the work schedule of participants reasonably, and better analyze and fix bugs. It can further clarify the roles of different participants in the bug fixing process, identify unreasonable behaviors of them, and ultimately standardize the bug fixing process.

Bugs with different priorities have different effects on software projects \citep{KoBaGo2016}. Understanding priorities of bugs can help developers solve bugs in a proper manner \citep{ZoLoChXiFeXu2018}. Therefore, it is important to assign and manage priorities of bugs appropriately. If the priorities of a substantial number of bugs are changed, it indicates delays in fixing critical bugs \citep{MeMa2008,ShBeChSi2013,ChKu2020,FeKhYiHa2012,KuSi2020}. Deepening the understanding of bug priority changes can help solve serious bugs as soon as possible and avoid delays, identify areas that need tool support to automatically verify bug priority change requests, and better record these changes \citep{AlFeKeSh2020}.

According to the data extracted from issue tracking systems, a proportion of bugs underwent priority changes after they were reported. The changed priorities of those bugs may negatively impact on the bug fixing process. Hence, it is valuable to understand in depth the phenomenon of bug priority changes. In this work, we conducted an empirical study to investigate bug priority changes through analyzing the history and comments of bugs in issue tracking systems and related commits to the bugs.

Due to certain reasons in software development, the priority of a bug may change \citep{AlFeKeSh2020,GoSo2021}, and the trends of change of priorities of different bugs may be different. For instance, due to time pressure, the scheduled bug fixing tasks may be delayed, and hence, the priorities of the involved bugs are changed to lower priority levels (e.g., from Critical to Minor). In addition, some bugs may be found to result in severer consequences than expected; thus, the priorities of such bugs will be changed to higher priority levels (e.g., from Minor to Critical). The time when the priority change happens can be different for different bugs. The priority of a bug may be changed before bug fixing is started, or when bug fixing is in progress, and it can also be changed after a closed or resolved bug is reopened. Finally, the personal habits and styles of bug participants may also have an impact on priority changes. For instance, a bug participant may prefer to allocate a lower priority to a bug than other bug participants do.

Currently, there lacks a comprehensive understanding on bug priority changes, which may negatively impact the bug fixing process.
To get a deep understanding on bug priority changes, we conducted an empirical study on 32 non-trivial Apache projects.
The main contributions are summarized as follows:
 
\begin{itemize}\setlength{\itemsep}{0pt}\setlength{\parskip}{0pt}
  \item To the best of our knowledge, this work is a first attempt to explore the phases and patterns of bug priority changes.
  \item Twenty-four patterns of bug priority changes are identified to characterize the process of bug priority changes. 
  \item We explored reasons for priority changes from the perspectives of the change complexity of bug-fixing commits and the communication complexity of bugs.
  \item We confirmed that the priority change is affected by several human factors.
\end{itemize}

The remaining of this paper is organized as follows. Section \ref{chap:relat} presents the related work; Section \ref{chap:case} describes the design of the empirical study; Section \ref{chap:study} presents the results of the study; Section \ref{chap:discus} discusses the study results; Section \ref{chap:threats} identifies the threats to validity of the results; and Section \ref{conclusions} concludes this work with future research directions.

\section{Related Work}\label{chap:relat}
Most existing works regarding bug priority focus on the influence of bug priority on the development process or the prediction of bug priority, and there is only one existing study on bug priority changes. Therefore, in Section \ref{chap:BugPriorityChange}, we compare our study with the only research on bug priority changes; in Section \ref{chap:BugPCImpact}, we discuss the related work on the role and impact of bug priority on the bug-fixing process; in Section \ref{chap:BugPCPrediction}, we present the recent studies on using machine learning or deep learning to predict bug priority.

\subsection{Bug Priority Change}
\label{chap:BugPriorityChange}
To our knowledge, only one existing work is focused on bug priority changes. Almhana et al. interviewed practitioners in industry and made quantitative analysis to understand the reasons why bug priorities are changed \citep{AlFeKeSh2020}. The reasons are summarized as follows: 
1)	the dependency of another bug’s fix incorrect priority, 
2)	type/domain of project, 
3)	category of the bug report, 
4)	lack of time/heavy workload/tight schedule, 
5)	accident, 
6)	hot-fix request, and
7)	business requirements.

Since the work of Almhana et al. \citep{AlFeKeSh2020} is the only one that investigates bug priority changes, we compare the study results of their work with our current work. First, to study when the priority will change, Almhana et al. answered this question from the perspective of stakeholders' work schedules and project version release time; in contrast, we answered this question from the perspective of the bug life cycle, which helps participants adjust their work focus and schedule based on the bug life cycle. Second, Almhana et al. made an in-depth investigation on the causes of bug priority changes; in comparison, we provided a detailed description of the characteristics of bug priority changes, which is valuable for practitioners in analyzing bug priority and fixing bugs. Third, when investigating the human factors of bug priority changes, Almhana et al. studied the modification of priorities by stakeholders from the perspective of stakeholder types, while we focused on exploring whether there are specific participants who modify bug priorities in a different manner from others, which helps identify unreasonable behaviors during the bug fixing process. Finally, our study is the first attempt to link bug priority changes with code commits and team communications, which helps to gain a deeper understanding of the reasons for bug priority changes. In summary, we studied bug priority changes from various perspectives, which fills the knowledge gap in this field and expands our understanding of bug priority changes.


\subsection{Impact of Bug Priority on the Bug Fixing}
\label{chap:BugPCImpact}
\textbf{Bug priority is used in some studies to predict bug-fixing time.} Akparinasaji et al. used bug priority to build the KNN model for predicting bug-fixing time \citep{AkCaBe2018}. Habayeb et al. used priority change as a factor to build a hidden Markov model to predict bug-fixing time \citep{HaMuMiBe2017}. Vicira et al. included bug priority and other data fields to build a dataset for bug report evolution, and trained three machine learning models to estimate the bug-fixing time \citep{ViMaRoGoPa2022}. Yuan et al. mentioned that although bug priority and repair strategy may affect repair time, their experimental results showed that they had no necessary correlation \citep{YuLuSuLi2020}.

\textbf{Some studies explored the impact of bug priority on bug-fixing process from different perspectives.} Gavidia-Caldcron et al. pointed out that when the priority of an issue (which can be a bug) is raised to a level higher than its real assessment, it will hinder software development, and they used Game Theory to understand and fix the problem \citep{GaSaHaBa2019}. Motwani et al. indicated that automated Java repair techniques are moderately more likely to produce patches for high-priority bugs, while automated C repair techniques are not correlated with bug priority \citep{MoSaJuBr2018}. Etemadi et al. used bug priority to develop a scheduling-driven approach to effectively assign bug repair tasks to developers \citep{EtBuAkRo2021}. 

\subsection{Bug Priority Prediction}
\label{chap:BugPCPrediction}
\textbf{Machine learning algorithms were widely used in the field of bug priority prediction.} Jaweria et al. proposed and evaluated a priority recommendation approach based on classifiers using Naive Bayes and Support Vector Machine (SVM) \citep{KaMa2012}. Alenezi et al. presented and evaluated an approach to predict the priority of a reported bug using Naive Bayes, Decision Trees, and Random Forest \citep{AlBa2014}. Sharma et al. used multiple machine learning techniques such as SVM, Naive Bayes, K-Nearest Neighbors, and Neural Networks. In their results of cross-project validation, the accuracy of prediction of bug priority is above 70\% except Naive Bayes \citep{ShBeChSi2013}. Yuan et al. proposed a method to predict bug priority by machine learning, which takes advantage of several factors such as temporal, textual, author, related-report, severity, and product \citep{YuLoXiSu2015}. 

\textbf{Neural networks and deep learning have also been used in bug priority prediction in recent years.} Yu et al. proposed neural network techniques to predict bug priority, adopted an evolutionary training process to solve error problems associated with new features, and reused datasets from similar software systems to accelerate the convergence of training \citep{YuTsZhWu2010}. Kumari and Singh built improved classifiers using Naive Bayes and deep learning and considered measures such as severity, summary weight, and entropy to predict bug priority \citep{KuSi2020}. 
Izadi et al. proposed a two-stage approach to predict the priority level after the opening of an issue using feature engineering methods and state-of-the-art text classifiers \citep{IzAkHe2022}.

\section{Study Design}\label{chap:case}
In order to investigate in depth the phenomenon of bug priority changes, we performed an empirical study on Apache Open Source Software (OSS) projects.
In this section, we describe the empirical study, which was designed and reported following the guidelines proposed by Runeson and H{\"o}st \citep{RuHo2009}.

\subsection{Objective and Research Questions}\label{DesignRQ}
The goal of this study, described using the Goal-Question-Metric (GQM) approach \citep{Ba1992}, is to \emph{analyze} the phenomenon of bug priority changes in depth, \emph{from the point of view of} software developers \emph{in the context of} OSS development. On the basis of the aforementioned goal, we have formulated five research questions (RQs), which are described as follows.

\begin{itemize}
\setlength{\itemsep}{0pt}
\setlength{\parsep}{0pt}
\item [\textbf{RQ1:}] \textbf{What is the proportion of bugs with priority changes?}\\
\textbf{Rationale:} With this RQ, we investigate the proportion of bugs with priority changes over all bugs of software projects, which gives practitioners and researchers a basic understanding on the state of bug priority changes. 

\item [\textbf{RQ2:}] \textbf{When is the bug priority changed?}\\
\textbf{Rationale:} With this RQ, we investigate the phases when the priorities of bugs are changed after they are reported. Understanding the time trend of bug priority changes can help a) practitioners arrange their work schedules reasonably, and b) researchers build automated detection tools to identify unreasonable priority changes.

\item [\textbf{RQ3:}] \textbf{How is the bug priority changed?}\\
\textbf{Rationale:} This RQ is focused on investigating the number, pattern, trend, and range of bug priority changes, and whether they will be affected by the priority itself. Investigating the process of bug priority changes can help practitioners optimize existing bug report documents, making it easier for practitioners to understand bug reports and obtain more information related to bugs, and researchers can use the characteristics exhibited by the priority change process to improve prediction models related to bugs.

\item [\textbf{RQ4:}] \textbf{Is there a significant difference between the complexity of bugs with priority changes and that of bugs without priority changes?}\\
\textbf{Rationale:} With this RQ, we further explore the relationship between the priority changes of bugs and their change complexity of bug-fixing commits and communication complexity. The results of this RQ may partially reveal the reasons for bug priority changes from the perspective of workload and team communication.

\item [\textbf{RQ5:}] \textbf{Do human factors play a role in bug priority changes?}\\
\noindent \textbf{Rationale:} With this RQ, we study whether the bug priority change is related to different types of priority modifiers, whether priorities reported by specific participants or priorities allocated by specific participants are more likely to be modified, and whether specific participants tend to modify bug priorities. The result of this RQ can help a) practitioners standardize the priority allocation process, and b) researchers further investigate the roles of different participants in the bug fixing process.

\end{itemize}

\subsection{Cases and Unit Analysis }\label{CasesandUnitAnalysis}
This study investigates multiple OSS projects, i.e., cases, and each bug and its corresponding bug-fixing commit is a single unit of analysis.

\subsection{Case Selection}\label{CaseSelection}
In this study, we only investigated Apache OSS projects. The reason is that the links between bugs and corresponding bug-fixing commits tend to be well recorded in the commit messages of those projects. For selecting each case (i.e., OSS project) included in our study, we applied the following criteria:
 \begin{itemize}\setlength{\itemsep}{0pt}\setlength{\parskip}{0pt}
  \item C1: The five bug priority levels, i.e., Blocker, Critical, Major, Minor, and Trivial, are adopted to label the priority of each bug in the project.
  \item C2: The age of the project is more than 5 years.
  \item C3: The number of revisions (i.e., commits) of code repository of the project is more than 3,000.
  \item C4: The number of bugs with priority changes in the project is more than 150.
\end{itemize}
Selection criterion C1 was set to ensure that the priority of each bug is explicitly defined. Criteria C2-C4 were set to ensure that the selected projects are non-trivial and the resulting dataset is big enough to be statistically analyzed. We selected all Apache projects (659 projects in total, and the initial list of projects have been made available in the replication package \citep{dataset}) and screened them, leaving only those that met the C1-C4 criteria, resulting in the final 32 projects. For example, \textit{Beam} is excluded by C1, \textit{Iceberg} is excluded by C2, \textit{Zipkin} is excluded by C3, and \textit{ZooKeeper} is excluded by C4.

\subsection{Data Collection}\label{DataCollection}

\subsubsection{Data Items to be Collected}
\begin{table}[]
\caption{Data items to be collected for each bug and their mapping to the RQs.}
 \centering
\scalebox{0.82}{
\begin{tabular}{p{0.04\columnwidth} p{0.23\columnwidth} p{0.65\columnwidth} p{0.07\columnwidth} }
\hline
\textbf{\#} & \textbf{Name} & \textbf{Description}                                                                                                               & \textbf{RQ}                                        \\ \hline
D1          & IsChanged  & Whether priority of the bug was changed.  & RQ1, RQ4                                                 \\ 
D2          & Priority & The final priority level of a bug. Five priority levels are defined in JIRA: Blocker, Critical, Major, Minor, and Trivial, from the highest to the lowest level.                                                   & RQ2, RQ3                                               \\ 
D3          & ChangeInterval & The time interval between when the bug was reported and when priority was changed for the first time. & RQ2
              \\ 
D4          & ChangePhase & The period of all priority changes of a bug. This item describes the periods of priority changes of a bug during its lifecycle.   & RQ2, RQ3  \\ 
D5          & PrioritySequence & The sequence of changed priority levels. This item describes the changes of the priority of a bug during its lifecycle. & RQ3 \\ 
D6          & ChangePattern & The pattern of priority changes. This item describes the priority change pattern followed by the bug.   & RQ3                                               \\ 
D7          & LOCM          & The number of lines of code modified to fix a bug. & RQ4                                                \\ 
D8          & NOFM          & The number of files (for Java) modified to a bug.  & RQ4
                                   \\ 
D9          & NOPM          & The number of packages (for Java) modified to a bug.  & RQ4                                                \\ 
D10         & Entropy          & The normalized entropy of the modified source files for fixing a bug during the last 60 days \citep{LiLiLiMoLi2020}.                                      & RQ4    
\\ 
D11          & NOC          & The number of comments on JIRA to a bug. & RQ4                                                \\ 
D12          & TLC          & The total length of all comments to a bug (in bytes).  & RQ4
                                   \\ 
D13          & NOCR         & The number of commenters to a bug.  & RQ4                                                \\ 
D14         & PriorityModifier  &The author of a priority change of a bug. There may be multiple PriorityModifiers for a bug since it may undergo multiple priority changes made by different practitioners.  & RQ5                                                \\ 
\hline
\end{tabular}
}
\label{table:dataitem}
\end{table}
To answer the five RQs, we took a bug and its corresponding bug-fixing commit as the unit of analysis and the data items to be collected are shown in Table \ref{table:dataitem}. 
Considering that data items D1 and D3 are straightforward, we only explain data items D2 and D4-D14 in detail.

\textbf{D2: Priority}. In JIRA, five priorities, Blocker, Critical, Major, Minor, and Trivial, are clearly defined \citep{ASF} as follows:
\begin{itemize}\setlength{\itemsep}{0pt}\setlength{\parskip}{0pt}
  \item \textbf{Blocker}: a time-sensitive issue that is hindering a basic function of a project.
  \item \textbf{Critical}: a time-sensitive issue that is disrupting the project, but does not hinder basic functions.
  \item \textbf{Major}: this issue needs attention soon, but is not hindering basic functions. Most requests for new resources fall into this category. 
  \item \textbf{Minor}: this issue needs attention, but is not time-sensitive and does not hinder basic functions.
  \item \textbf{Trivial}: this issue is minimal and has no time constraints.
\end{itemize}


\begin{figure*}
\centerline{\includegraphics[width=6.0in]{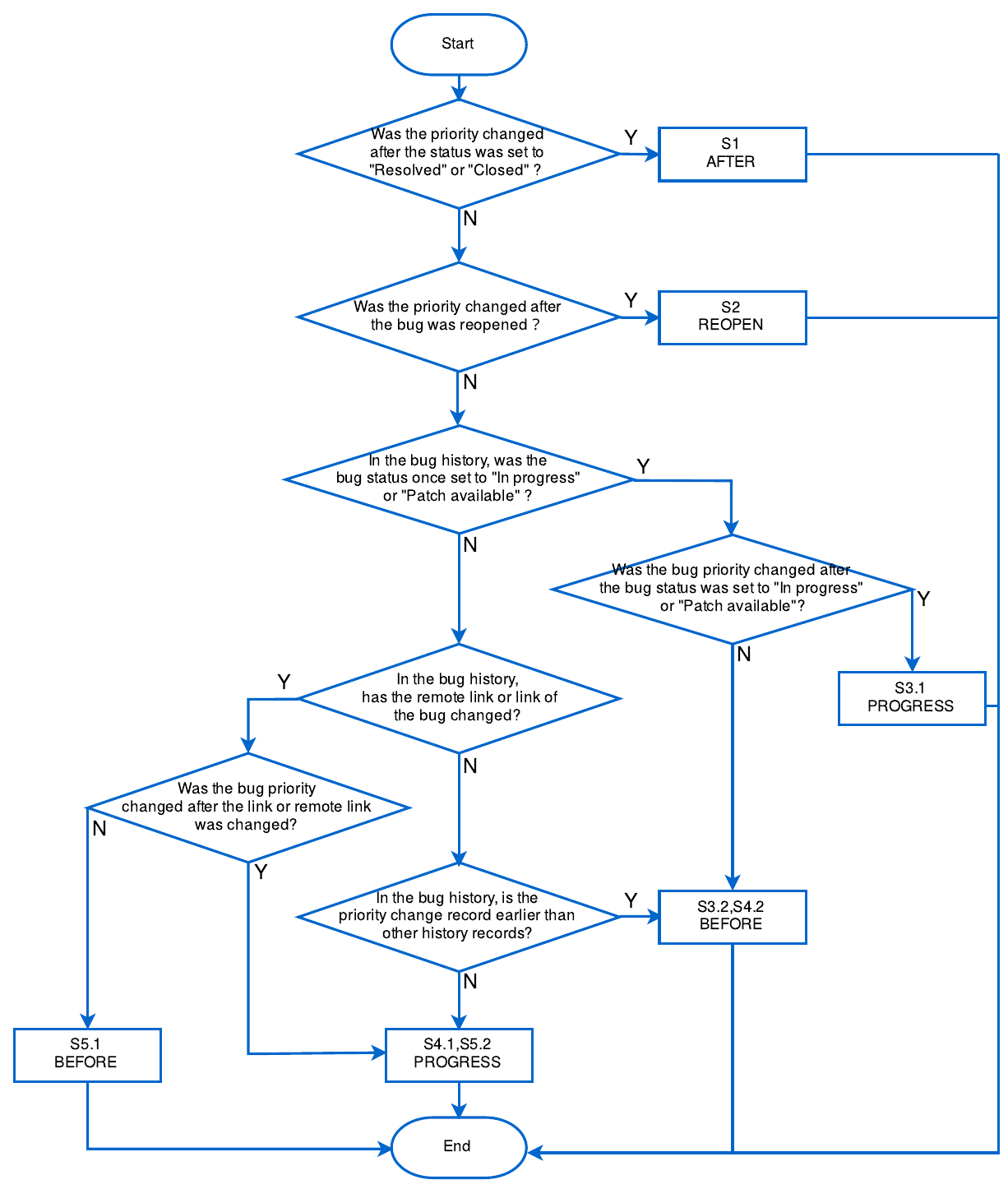}}
\caption{Procedure of bug priority change phase identification.}
\label{fig:changephaseiden}
\end{figure*}

\begin{figure*}
\centerline{\includegraphics[width=7.0in]{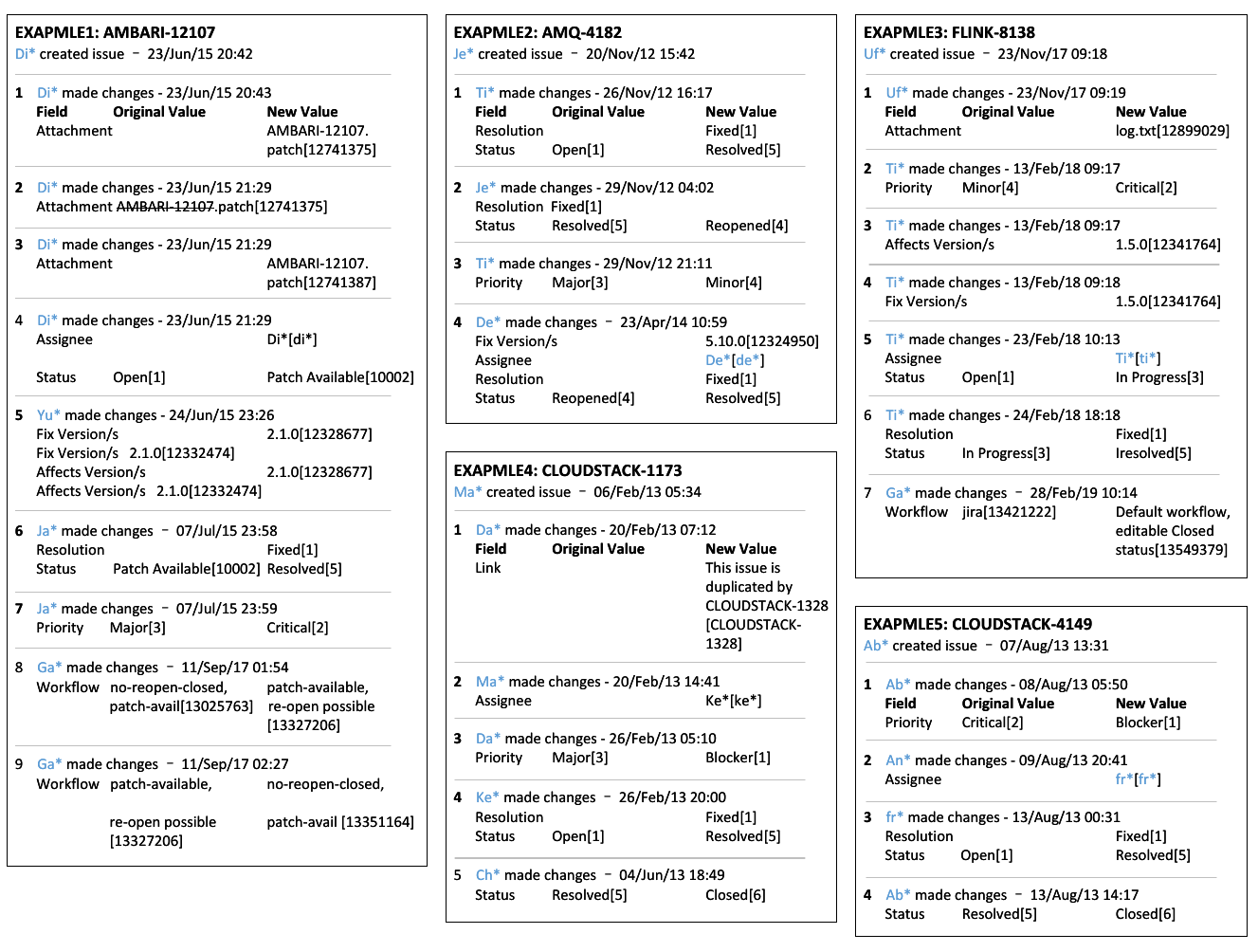}}
\caption{Examples of historical information of five bugs.}
\label{fig:5example}
\end{figure*}

\textbf{D4: ChangePhase}. We explain data item D4 (i.e., ChangePhase) in detail, since the task of collecting D4 is relatively complicated. We divide the period when the priority of a bug was changed into 4 phases: 
(1) \textbf{BEFORE}, if the priority of a bug was changed before the bug was handled; 
(2) \textbf{PROGRESS}, if the priority of a bug was changed during the process of handling the bug; 
(3) \textbf{REOPEN}, if the priority of a bug was changed after the bug was reopened; 
and (4) \textbf{AFTER}, if the priority of a bug was changed after the status of the bug was turned into “Resolved” or “Closed”. 
Because the priority of a bug may be changed multiple times, we investigated the history of each bug to identify the change phase of each priority change.
The procedure of labeling the ChangePhase of each bug priority change is defined as follows (also shown in Figure \ref{fig:changephaseiden}):

\begin{itemize}\setlength{\itemsep}{0pt}\setlength{\parskip}{0pt}
    \item \textbf{S1}: Check whether the priority was changed after the status of the bug was changed to  ``Close'' or ``Resolve''. If so, the ChangePhase of this priority change is labeled as AFTER.
    \item \textbf{S2}: Check whether the priority of the bug was changed after the bug was reopened. If so, the ChangePhase of this priority change is labeled as REOPEN.
    \item \textbf{S3}: If the status of the bug was changed to ``In progress'' or ``Patch available'' in the history of the bug, check whether priority was changed after bug status was changed to ``In progress'' or ``Patch available''.
    \begin{itemize}
        \item \textbf{S3.1}: If so, the ChangePhase of this bug priority change is labeled as PROGRESS.
        \item \textbf{S3.2}: If not, the ChangePhase of this priority change is labeled as BEFORE.
    \end{itemize}
    \item \textbf{S4}: If the bug status has not changed to ``In Progress'' or ``Patch Available'' in the history of the bug, but its remote link or link has changed, check whether the priority of the bug was changed after its remote link or link was changed.
    \begin{itemize}
        \item \textbf{S4.1}: If so, the ChangePhase of this priority change is labeled as PROGRESS.
        \item \textbf{S4.2}: If not, the ChangePhase of this priority change is labeled as BEFORE.
    \end{itemize}
     \item \textbf{S5}: If the bug status has not changed to ``In Progress'' or ``Patch Available'' in history of bug, and its remote link or link has not changed, check whether the priority change record is earlier than other history records.
    \begin{itemize}
        \item \textbf{S5.1}: If so, the ChangePhase of this priority change is labeled as BEFORE.
        \item \textbf{S5.2}: If not, the ChangePhase of this priority change is labeled as PROGRESS.
    \end{itemize}
    
\end{itemize}

For each bug, we collected all its historical change items and arranged them in a chronological order (we provided five examples of historical information for five bugs in Figure \ref{fig:5example}), and the historical information of a bug consists of multiple change records, each of which consists of the \textit{Modifier}, \textit{Change Time}, \textit{Field Name}, \textit{Original Value}, and \textit{New Value}. Then we calculated the ChangePhase for this bug priority change according to the above process (i.e., situation S1 to situation S5). The division of situation S1 and situation S2 is easy to understand (corresponding examples are shown in EXAMPLE1 and EXAMPLE2 in Figure \ref{fig:5example}), and here we explain the reasons for dividing situation S3, situation S4, and situation S5. For some bugs, their historical status is displayed as ``In progress'' or ``Patch available'', indicating that such bugs have started to be handled. For example, the status of EXAMPLE3 in Figure \ref{fig:5example} was changed from ``Open'' to ``In progress'' at 10:13 on 23/Feb/18 (corresponding to change number 5). Prior to  this status change, a priority change occurred at 09:17 on 13/Feb/18 (corresponding to change number 2). Therefore, we believe that this priority change corresponds to situation S3.2, which is ``BEFORE''. However, in the life cycle of some bugs, their status does not change to ``In progress'' or ``Patch available'', as the process of adjusting the bug status is not very rigorous. For example, the status of EXAMPLE4 and EXAMPLE5 in Figure \ref{fig:5example} were directly change from ``Open'' to ``Resolved''. One of the reasons is that the bug (e.g., EXAMPLE4 in Figure \ref{fig:5example}) is linked to another bug, such as a duplicate bug. After fixing the linked bug, this bug is also closed. Therefore, if the status of a bug is not changed to ``In progress'' or ``Patch available'' but is linked to another bug, we believe that the bug begins to be handled when it is linked to another bug, such as EXAMPLE4 in Figure \ref{fig:5example}. The priority change of the bug occurs after it is linked to another bug, so this priority change corresponds to situation S4.1, which is ``PROGRESS''. If the ChangePhase for a priority change cannot be determined after the above process, then we can only compare the priority change with other historical change records. We believe that in this case, as long as any field of a bug (except for ``Priority'') changes, it can be considered that the bug has started to be handled. For example, EXAMPLE5 in Figure \ref{fig:5example}, the priority change occurred before any field was modified, corresponding to situation S5.1, which is ``BEFORE''.


\textbf{D5: PrioritySequence.} Different priority levels correspond to different numbers. Bug priority levels Blocker, Critical, Major, Minor, and Trivial in JIRA correspond to 1, 2, 3, 4, and 5 respectively. The PrioritySequence of a bug is recorded by a sequence of priority numbers in chronological order of priority changes. For instance, when the priority of a bug was changed from Blocker to Minor, and then from Minor to Major, the PrioritySequence is recorded as ``143''.

\begin{figure*}    \centering{\includegraphics[width=6.3in]{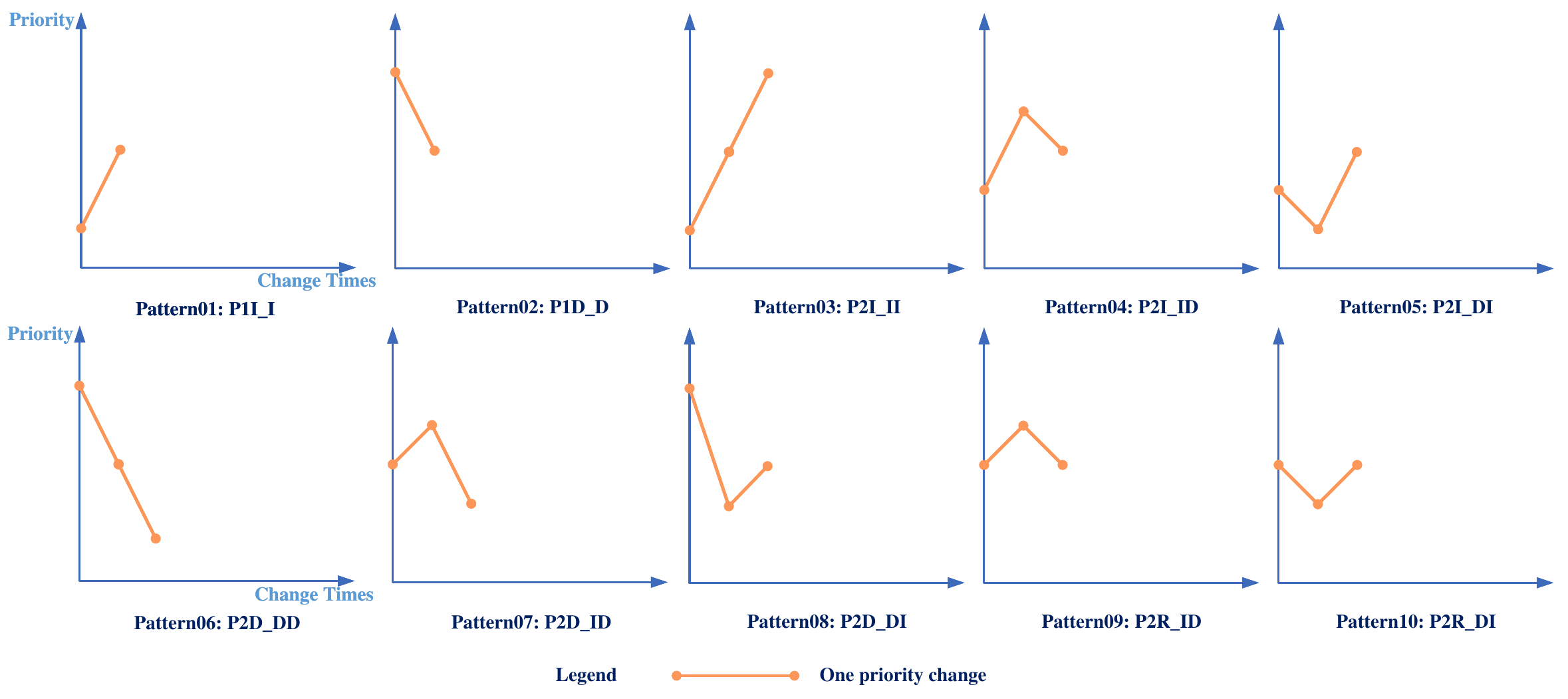}}
    \caption{Bug priority change patterns when a bug's priority is changed for one or two times.}
    \label{fig:Patterns12}
\end{figure*}

\begin{figure*}    \centering{\includegraphics[width=6.3in]{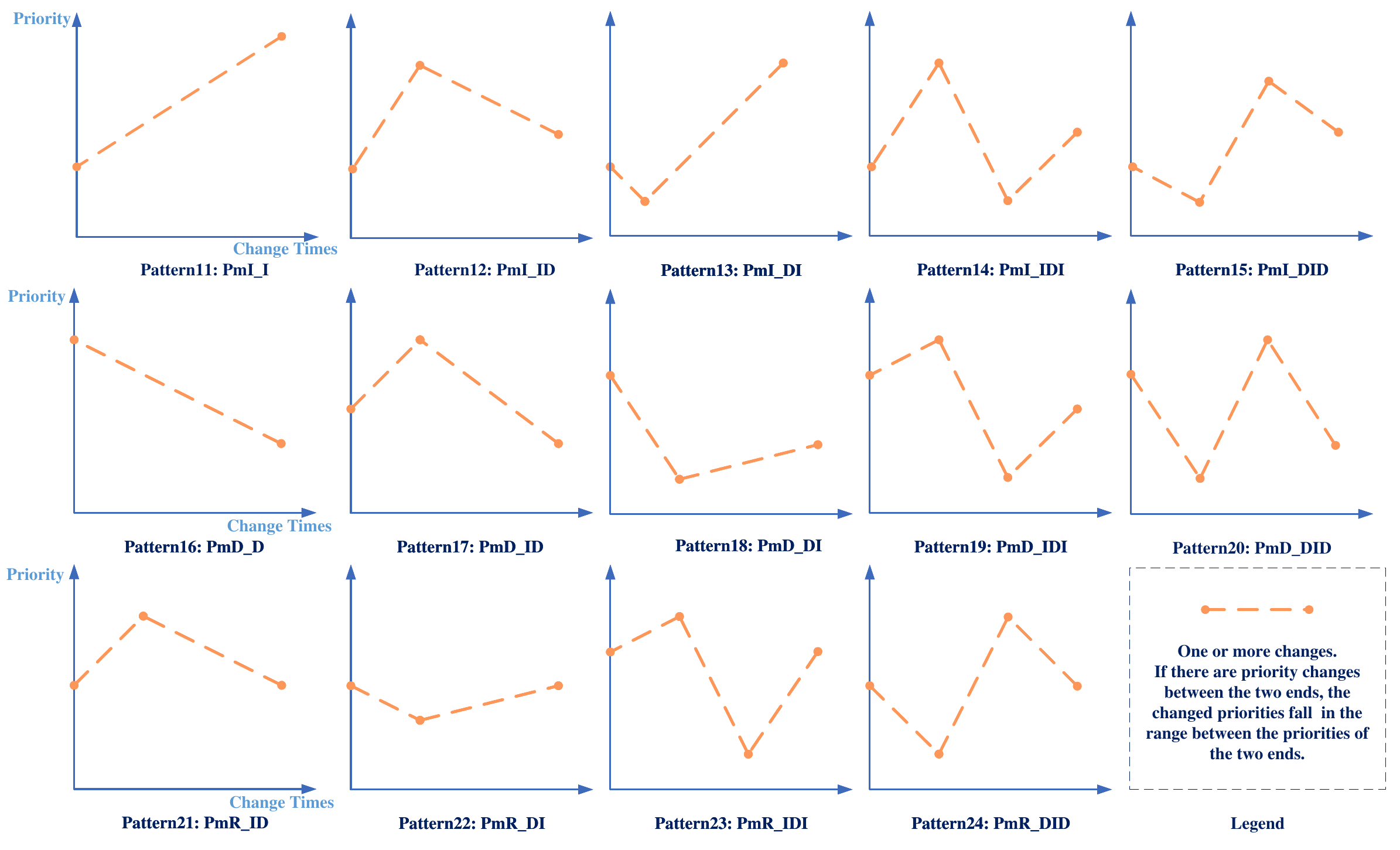}}
    \caption{Bug priority change patterns when a bug's priority is changed for three or more times.}
    \label{fig:Patterns3Plus}
\end{figure*}

\textbf{D6: ChangePattern.} To investigate how the bug priority is changed, we summarize all possible ways of bug priority changes based on a 4-point model (the four points are initial priority, final priority, highest priority, and lowest priority of a bug) into a list of bug priority change patterns (a total of 24 change patterns), which are shown in Figure \ref{fig:Patterns12} and Figure \ref{fig:Patterns3Plus}. Compared with only considering the initial priority and final priority, we use four points to summarize all patterns of priority changes, which can reflect the priority change process in more detail, rather than just telling the final result of the priority changes. When we summarize the priority change patterns, we distinguish different number of times the priority changes. This is because the vast majority of bugs only undergo one or two priority changes, and the process of one or two priority changes is simpler. Therefore, we list all points in the process of one or two priority changes, and every two neighboring points are connected by a solid line in Figure \ref{fig:Patterns12}. Correspondingly, only a very small number of bugs undergo three or more priority changes, and the change process is relatively complicated. We use four points to describe the change process, and every two neighboring points are connected by a dotted line in Figure \ref{fig:Patterns3Plus}. We have arranged a sequence number for each pattern and assigned an alias based on its changing characteristics. The alias for a pattern is in one of the following forms: PnC\_X, PnC\_XY, PnC\_XYZ. The rules for taking aliases are described as follows: (1) \textit{P} is the abbreviation of word ``Pattern''; (2) letter \textit{n} following letter \textit{P} represents the number of priority changes for the bug: 1, 2, and \textit{m} denote that the bug undergoes 1, 2, and more than 2 priority changes respectively; (3) letter \textit{C} represents the final result of the priority changes for the bug: \textit{I} indicates that the bug priority is changed to a higher level, \textit{D} indicates that the bug priority is changed to a lower level, and \textit{R} indicates that the bug priority returns to the original level; and (4) the string of letters after the underscore represents the process of bug priority changes. 


Pattern01 $\sim$ Pattern02 and Pattern03 $\sim$ Pattern10 in Figure \ref{fig:Patterns12} show all cases in which the bug priority changes for one or two times, respectively. Specifically, the alias of Pattern01 is P1I\_I, and the fact that a bug's priority change pattern is P1I\_I means that: this bug undergoes a priority change, its final priority is changed to a higher level, and the change process is that the priority is increased once. P1D\_D can be interpreted in a similar way. In P2I\_II, P2I\_ID, and P2I\_DI, after different change processes, the bug priority is changed to a higher level finally; in P2D\_DD, P2D\_ID, and P2D\_DI, the bug priority is changed to a lower level finally; in P2R\_ID and P2R\_DI, the bug priority returns to the original level finally after a change to a higher or lower level.

In Figure \ref{fig:Patterns3Plus}, all patterns show the cases in which the bug priority is changed three or more times. We have summarized 14 patterns according to the initial priority, the final priority, the highest priority, and the lowest priority of a bug. In Pattern11 $\sim$ Pattern15 (i.e., PmI\_I, PmI\_ID, PmI\_DI, PmI\_IDI, PmI\_DID), after ups and downs the bug priority is changed to a higher level finally. In Pattern16 $\sim$ Pattern20 (i.e., PmD\_D, PmD\_ID, PmD\_DI, PmD\_IDI, PmD\_DID), after ups and downs the bug priority is changed to a lower level finally. In Pattern21 $\sim$ Pattern24 (i.e., PmR\_ID, PmR\_DI, PmR\_IDI, PmR\_DID), after ups and downs the bug priority returns to the original level. Each bug with priority changes can be labeled as one of the 24 bug priority change patterns.


\textbf{D7 $\sim$ D13: LOCM, NOFM, NOPM, Entropy, NOC, TLC and NOCR.} We chose D7 $\sim$ D10 as indicators for change complexity of bug-fixing commits, as these indicators have been widely used in previous studies \citep{OlFeStCrCoGa2020,NaRiSh2019,AlStHe2022,LiLiLiMoLi2020}. D11 $\sim$ D13 were chosen as indicators of the communication complexity of a bug since these three indicators are directly obtainable information, and they are simple, easy to understand, and can intuitively represent the communication complexity of a bug. D7 $\sim$ D9 and D11 $\sim$ D13 are clearly defined, and here we only explain the definition of the entropy of the modified source files in a commit (i.e., D10) in detail~\citep{Ha2009}. Suppose that the modified source files of commit $c$ are $\{f_1,f_2,\cdots,f_n\}$, and file $f_i\left(1\leq i\leq n\right)$ has been modified $m_i$ times (i.e., in $m_i$ commits) during a period of time before this commit.
Let 
\begin{equation}
     p_i = m_i/\sum_{i=1}^nm_i.
\end{equation} 
Then, the entropy 
\begin{equation}
    H(n) = -\sum_{i=1}^np_ilog_2 p_i. 
\end{equation} 
The normalized entropy 
\begin{equation}
\tilde{H}(n) =
\begin{cases}
H(n)/log_2n& \text{n \textgreater 1,}\\
0& \text{n = 1.}
\end{cases} 
\end{equation}
In this study, the period is set to 60 days (including the day when commit $c$ happened), which is chosen according to the period set by Hassan and Li et al. when calculating the entropy of the modified source files for fixing a bug \citep{Ha2009, LiLiLiMoLi2020}.
 
\textbf{D14: PriorityModifier.} People related to a bug are referred to as participants. We divide participants into five types: Reporter, Assignee, Commenter, FieldModifier, and Other. PriorityModifier can be these five types of participants. Reporter denotes the participant who reports the bug; Assignee denotes the participant to whom the bug is assigned; Commenter denotes the participant who makes comments on this bug; FieldModifier denotes the participant who modifies fields of the bug except for the field of priority; Other denotes the participant who does not preform any of the above activities.

\subsubsection{Data Collection Procedure}\label{datacollectionprocedure}
We have formulated the following rules in advance to filter and process the priority change records that we have collected.
\begin{itemize}\setlength{\itemsep}{0pt}\setlength{\parskip}{0pt}
    \item \textbf{R1}: Check whether the original priority of the bug or the priority of the bug was changed is null. If so, this situation is not considered as a priority change of the bug.
    \item \textbf{R2}: Check whether the priority of the bug was changed by the same PriorityModifier within 5 minutes after the bug was reported. If so, this situation is not considered as a priority change, but the priority after change is considered as the initial priority. For example, when a bug is created, the priority is Major. One minute later, the same PriorityModifier changes the priority to Critical. Then we do not consider this priority modification as a priority change, and treat the priority of this bug when created as Critical.
    \item \textbf{R3}: Check whether the bug priority is changed back to the original priority by the same PriorityModifier within 5 minutes after the priority is modified. If so, these two priority modifications do not count.
    \item \textbf{R4}: Check whether the priority of the bug is modified by the same PriorityModifier within 5 minutes after the priority is modified, but it is not changed back to the original priority. If so, the two priority changes are combined into one, that is, from the original priority to the final priority. The time of change is subject to the time of the second change. For example, a PriorityModifier changed the priority of a bug from Major to Critical at 18:00 on October 26, 2020, and the same PriorityModifier changed the priority from Critical to Blocker at 18:01 on October 26, 2020. Then we merge the two priority modification records, that is, this PriorityModifier changed the bug priority from Major to Blocker at 18:01 on October 26, 2020.
\end{itemize}

\textbf{R1} is formulated to filter the situation where the priority is null. \textbf{R2}, \textbf{R3}, and \textbf{R4} are formulated to deal with the problem of multiple modifications by the same PriorityModifier in a short time for personal reasons, such as regret. This time period is set as 5 minutes. The reason is that we manually checked the multiple modifications in a short period of time, and found that the modifications after the creation of a bug or the modification of the priority are mainly made in the next 5 minutes, while the modifications after 5 minutes are scattered.

\begin{figure}
\centerline{\includegraphics[width=2.4in]{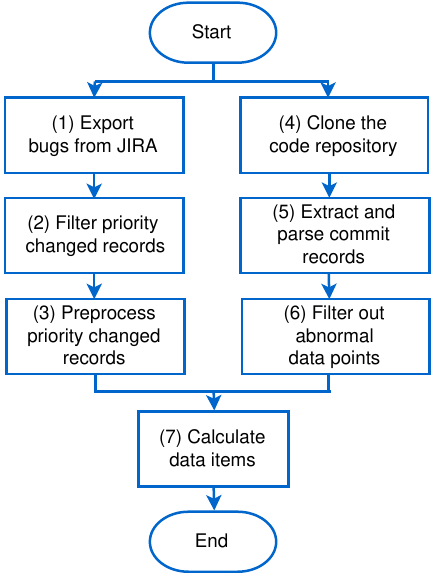}}
\caption{Procedure of data collection.}
\label{fig:proceduredatacollection}
\end{figure}

The procedure of collecting the data items listed in Table \ref{table:dataitem} consists of seven steps, as shown in Figure \ref{fig:proceduredatacollection}. For each selected project, the details of the steps are described as follows.

\begin{itemize}\setlength{\itemsep}{0pt}\setlength{\parskip}{0pt}

\item \textbf{Step 1}: Extract bugs from JIRA. According to the APIs\footnote{\href{https://developer.atlassian.com/cloud/jira/platform/rest/v3/intro/}{https://developer.atlassian.com/cloud/jira/platform/rest/v3/intro/}} provided by JIRA, we developed a tool to extract issues. We used this tool to obtain all issues of Apache OSS Projects.
\item \textbf{Step 2}: Filter out priority change records. According to R1, we filtered out the records with null priority.
\item \textbf{Step 3}: Preprocess priority change records. According to R2-R4, we processed multiple priority changes made by the same PriorityModifier in a short time.
\item \textbf{Step 4}: Clone the code repository. With the TortoiseGit tool, we cloned the Git repository of the project.
\item \textbf{Step 5}: Extract and parse commit records. First, we removed the duplicate commit records introduced for unknown reasons and only retained one of the duplicate commit records. We found that there may be duplicate commit records in which all the change details are identical except for the revision ID. Second, we extracted bug ID from the message in each commit record and mapped each bug and commit record according to bug ID. Because not all the bugs in issues can be mapped to commits, the bugs after parsing the commits (i.e., the bugs that can be mapped to commits) are less than bugs in JIRA.
\item \textbf{Step 6}:  Filter out abnormal data points. We found that some bugs which fixing involves either more than 100 modified source files or over 10,000 lines of modified code \citep{LiLiLiMoLi2020}. These data items can affect the validity of our conclusions.
\item \textbf{Step 7}: Calculate data items. We counted the priority change records of each bug and calculated D1-D5. By analyzing the PrioritySequence (i.e., D5) of the bug, we can calculate the ChangePattern (i.e., D6) of each bug with priority changes. D7-D10 were calculated based on related commit records. D11 is an inherent field of each bug, which can be directly obtained. Finally, we collected PriorityModifier (i.e., D12) of each bug.

\end{itemize}

\subsection{Data Analysis}
The answers to RQ1, RQ2, and part of RQ3 can be obtained by descriptive statistics. Here, we only describe the calculation process of bug priority change range for RQ3, and details of data analysis for RQ4 and RQ5.

\textbf{Priority change range.} To answer RQ3 about the priority change range, we calculate the average priority change range of the bug according to the PrioritySequence of each bug. For example, if the PrioritySequence of a bug is 143, the priority change range is $(|4-1|+|3-4|)/2=2$.

\textbf{Change complexity of bug-fixing commits and communication complexity.} The change complexity of bug-fixing commits of a bug is indicated by LOCM, NOFM, NOPM, and Entropy. Communication complexity is indicated by NOC, TLC, and NOCR. To answer RQ4, in addition to descriptive statistics, we performed Mann-Whitney U tests \citep{Fi2013} to examine if two groups are significantly different from each other. Bugs are divided into two types: bugs with priority changes and bugs without priority changes. We calculated the indicators of change complexity of bug-fixing commits and the indicators of communication complexity for these two types of bugs, and then used Mann-Whitney tests to examine whether there is significant difference between the two types of bugs with respect to each indicator.
Since the variables to be tested do not necessarily follow a specific distribution, it is reasonable to use the Mann-Whitney U test -- a non-parametric test -- in this study. The test is significant at \emph{p-value} \textless{} 0.05, which means that the tested groups have a significant difference. In addition, we also calculated the effect size (\textit{Cohen'd}) to show the difference in size between the two sets of data.

\textbf{Types of PriorityModifier.} A PriorityModifier may belong to multiple types, for example, a PriorityModifier may be both the Reporter and Assignee of a bug, and has also made comments on the bug. We calculated the proportion and distribution of different types of PriorityModifier respectively. When calculating the proportion of different types of PriorityModifier, each PriorityModifier can belong to multiple types if applicable. 
When calculating the distribution of different types of PriorityModifier, each PriorityModifier only belongs to one type. The meaning of PriorityModifier type here changed, and two new PriorityModifier types were added. Their meanings are described as follows: Reporter denotes the person who reported the bug but is not the one assigned to the bug; the meaning of Assignee is opposite to that of Reporter; Reporter\&Assignee denotes the reporter and assignee of the bug; Commenter denotes the person who only makes comments when she is not a Reporter or Assignee; FieldModifier denotes the participant who only modifies fields of the bug except for the Priority field when the participant is not a Reporter or Assignee; Commenter\&FieldModifier denotes the person who makes comments and modifies fields of the bug except for the Priority field when she is not a Reporter or Assignee; Other denotes the participant who did not perform any of the above activities.

\textbf{Bug participants prone to making priority changes.} To study the possible human factors in bug priority changes, we investigated three specific groups of bug
participants: (1) participants whose reporting bug priorities are likely to be modified, denoted as PRs; (2) participants whose allocating bug priorities are likely to be modified, denoted as PAs; and (3) participants who are likely to modify bug priorities, denoted as PMs. First of all, according to RQ1, we can get the overall bug priority change probability of all projects (denoted as $p$). Then, we calculate the average number of bugs reported (denoted as $N_r$), average number of priorities allocated (i.e., the number of priorities modified, denoted as $N_a$), and average number of bugs involved by each core participant (denoted as $N_c$) in all projects. According to the work of \citep{ChLiLiZhLi2017}, supposing that the participants are ranked by the number of their involved bugs in a descending order in each project, core participants are the top participants whose total number of involved bugs reaches 80\% of the total number of bugs in the project. 
Definitions of PR, PA, and PM are described as follows: 1) If a participant reports at least $N_r$ bugs, and these bugs have their priority modified later with the probability of at least $2p$, then the participant is a PR; 2) If there are at least $N_a$ priorities allocated to a participant and these priorities will be modified with the probability of at least $2p$, then the participant is a PA; 3) If the ratio between the number of priorities modified by a participant and the number of bugs she involves in is at least $2p$, then the participant is a PM. The reason why we set the priority change probability at $2p$ here is that we regard that more than twice the overall priority change probability (i.e., $p$) is a higher priority change probability, while $N_r$, $N_a$, and $N_c$ are to ensure sufficient bug samples for each participant.

\begin{table*}[]
\centering
\caption{Demographic information of selected Apache OSS projects.}
\label{table:demographic}
\scalebox{0.81}{
\begin{tabular}{lrllrrr}
\hline
    \textbf{Project} & \multicolumn{1}{c}{\textbf{Age(y)}} & \textbf{Language}     & \textbf{\#Domain}         & \textbf{\#Revision} & \textbf{\#Committer} & \textbf{\#Bug in JIRA} \\ \hline
ActiveMQ         & 17                                  & Java                          & Message Middleware                     & 10,728               & 166                   & 5,058                   \\
Ambari           & 11                                  & Java, JavaScript, Python      & Hadoop Clusters Administration Tool                     & 24,110               & 236                   & 17,961                  \\
Arrow            & 8                                   & C++, Java, Go, Python         & Columnar Memory Format                     & 12,327               & 948                   & 5,576                   \\
Axis2            & 18                                  & Java, C, C++                  & Web Service Container                     & 14,030               & 87                    & 4,558                   \\
Camel            & 16                                  & Java                          & Integration Framework                     & 55,632               & 1,100                  & 5,758                   \\
CloudStack       & 12                                  & Java, Python                  & Infrastructure-as-a-Service Platform                     & 32,759               & 608                   & 7,861                   \\
Cordova          & 14                                  & JavaScript, Java, Objective-C & Mobile App Development Platform                     & 33,010               & 1,223                  & 8,780                   \\
Drill            & 10                                  & Java                          & SQL Query Engine                     & 4,313                & 218                   & 5,437                   \\
Flink            & 12                                  & Java, Scala                   & Distributed Processing Engine                     & 31,286               & 1,598                  & 11,003                  \\
Geode            & 12                                  & Java                          & Data Management Platform                     & 10,803               & 204                   & 5,429                   \\
Groovy           & 19                                  & Java, Groovy                  & Dynamic Language                     & 18,983               & 440                   & 6,921                   \\
Guacamole        & 12                                  & C, Java, JavaScript           & Clientless Remote Desktop Gateway                     & 8,069                & 141                   & 856                    \\
Hadoop           & 13                                  & Java                          & Distributed Processing Framework                     & 25,738               & 645                   & 23,798                  \\
HBase            & 16                                  & Java                          & Distributed Database                     & 19,428               & 600                   & 12,227                  \\
Hive             & 14                                  & Java                          & Data Warehouse                     & 16,040               & 487                   & 14,306                  \\
Hudi             & 6                                   & Java, Scala                   & Data Lake                     & 3,314                & 381                   & 1,177                   \\
Ignite           & 9                                   & Java, C\#, C++                & Distributed In-Memory Computing Platform                     & 22,146               & 456                   & 7,242                   \\
Impala           & 11                                  & C++, Java, Python             & Distributed SQL Query Engine                     & 10,379               & 243                   & 6,238                   \\
Jackrabbit-Oak   & 11                                  & Java                          & Hierarchical Content Repository                     & 17,839               & 77                    & 3,726                   \\
Kafka            & 11                                  & Java, Scala                   & Streaming Message Middleware                     & 10,361               & 1,116                  & 6,808                   \\
Lucene           & 21                                  & Java                          & Full-text Search Library                     & 35,469               & 386                   & 4,201                   \\
Mesos            & 12                                  & C++                           & Cluster Manager                     & 18,175               & 388                   & 4,922                   \\
NetBeans         & 9                                   & Java                          & Integrated Development Environment                     & 5,320                & 235                   & 4,740                   \\
NiFi             & 8                                   & Java                          & Processing Engine                     & 7,303                & 570                   & 4,427                   \\
OFBiz            & 16                                  & Java, JavaScript              & Java Web Framework                     & 32,911               & 89                    & 4,740                   \\
Ozone            & 5                                   & Java                          & Distributed Key-Value Store                     & 4,830                & 193                   & 2,471                   \\
Qpid             & 16                                  & Java, C++                     & Message Queue                     & 45,216               & 121                   & 5,097                   \\
Solr             & 21                                  & Java                          & Enterprise Search Platform                     & 32,002               & 228                   & 7,209                   \\
Spark            & 13                                  & Scala, Python, Java           & Distributed Processing Engine                     & 32,561               & 2,572                  & 15,508                  \\
Thrift           & 16                                  & C++, Java, C                  & Cross-Language Services Development                     & 6,553                & 562                   & 3,058                   \\
Traffic Server   & 13                                  & C++, Python, C                & Caching Proxy Server                     & 13,643               & 381                   & 3,026                   \\
Wicket           & 18                                  & Java                          & Java Web Framework                     & 21,108               & 142                   & 4,193                   \\ \hline
\end{tabular}%
}
\end{table*}

\section{Study Results}\label{chap:study}

We collected data items described in Table \ref{table:dataitem} from 32 non-trivial Apache OSS projects that were selected following the criteria set defined in Section \ref{CaseSelection}. The data of the selected projects were collected around the beginning of September 2022. The
replication package of this study has been made available online \citep{dataset}, including the raw data, code, calculation results, and a README file. The demographic information of the 32 projects is shown in Table \ref{table:demographic}. The age of each project falls in the range from 5 to 21 years, the number of revisions (i.e., commits) of each project falls in the range between 3,314 and 55,632, the number of committers of each project falls in the range between 77 and 2,572, and the number of bugs reported in JIRA falls in the range between 856 and 23,798. In the rest of this section, we present the results for each RQ.

\subsection{Proportion of Bugs with Priority Changes (RQ1)}

Table \ref{table:proportion} shows the proportion of bugs with priority changes for each selected project and that for all selected projects as a whole. In this table, \#BugPC denotes the number of bugs with priority change(s). Specifically, the proportion of bugs with priority changes for each selected project ranges from 1.7\% (project \emph{Ambari}) to 24.4\% (project \emph{Guacamole}), and when taking all projects as a whole, the proportion of bugs with priority changes is 8.3\%, which is relatively low. 



\begin{table}[]
\centering
\caption{Proportion of bugs with priority changes (RQ1).}
\label{table:proportion}
\scalebox{0.68}{
\begin{tabular}{lrrr}
\hline
\textbf{Project}      & \multicolumn{1}{c}{\textbf{\#Bug in JIRA}} & \multicolumn{1}{c}{\textbf{\#BugPC}} & \multicolumn{1}{c}{\textbf{\%}} \\ \hline
ActiveMQ              & 5,058                                       & 216                                                     & 4.3\%                             \\
Ambari                & 17,961                                      & 307                                                     & 1.7\%                             \\
Arrow                 & 5,576                                       & 290                                                     & 5.2\%                             \\
Axis2                 & 4,554                                       & 425                                                     & 9.3\%                             \\
Camel                 & 5,758                                       & 807                                                     & 14.0\%                            \\
CloudStack            & 7,861                                       & 870                                                     & 11.1\%                            \\
Cordova               & 8,780                                       & 813                                                     & 9.3\%                             \\
Drill                 & 5,437                                       & 579                                                     & 10.6\%                            \\
Flink                 & 10,296                                      & 2,049                                                   & 19.9\%                            \\
Geode                 & 5,429                                       & 150                                                     & 2.8\%                             \\
Groovy                & 6,918                                       & 352                                                     & 5.1\%                             \\
Guacamole             & 856                                         & 209                                                     & 24.4\%                            \\
Hadoop                & 23,798                                      & 2,244                                                   & 9.4\%                             \\
HBase                 & 12,227                                      & 1,142                                                   & 9.3\%                             \\
Hive                  & 14,306                                      & 509                                                     & 3.6\%                             \\
Hudi                  & 1,177                                       & 184                                                     & 15.6\%                            \\
Ignite                & 7,242                                       & 582                                                     & 8.0\%                             \\
Impala                & 6,238                                       & 1,431                                                   & 22.9\%                            \\
Jackrabbit-Oak        & 3,726                                       & 168                                                     & 4.5\%                             \\
Kafka                 & 6,808                                       & 617                                                     & 9.1\%                             \\
Lucene                & 4,201                                       & 201                                                     & 4.8\%                             \\
Mesos                 & 4,922                                       & 427                                                     & 8.7\%                             \\
NetBeans              & 4,740                                       & 314                                                     & 6.6\%                             \\
NiFi                  & 4,427                                       & 221                                                     & 5.0\%                             \\
OFBiz                 & 4,740                                       & 215                                                     & 4.5\%                             \\
Ozone                 & 2,471                                       & 194                                                     & 7.9\%                             \\
Qpid                  & 5,097                                       & 214                                                     & 4.2\%                             \\
Solr                  & 7,209                                       & 377                                                     & 5.2\%                             \\
Spark                 & 15,265                                      & 1,773                                                   & 11.6\%                            \\
Thrift                & 3,058                                       & 183                                                     & 6.0\%                             \\
Traffic Server        & 3,026                                       & 165                                                     & 5.5\%                             \\
Wicket                & 4,193                                       & 212                                                     & 5.1\%                             \\
\textbf{All projects} & \textbf{223,355}                            & \textbf{18,440}                                         & \textbf{8.3\%}                    \\ \hline
\end{tabular}
}
\end{table}

\begin{table}[]
\centering
\caption{Time intervals (in days) of bug priority changes (RQ2).}
\label{table:ChangeTimeInterval}
\scalebox{0.7}{
\begin{tabular}{lrrrrr}
\hline
\multirow{2}{*}{\textbf{Project}} & \multicolumn{1}{c}{\multirow{2}{*}{\textbf{Min}}} & \multicolumn{1}{c}{\multirow{2}{*}{\textbf{Median}}} & \multicolumn{1}{c}{\multirow{2}{*}{\textbf{Average}}} & \multicolumn{1}{c}{\multirow{2}{*}{\textbf{Max}}} & \multicolumn{1}{c}{\textbf{365 days}} \\ \cline{6-6} 
                                  & \multicolumn{1}{c}{}                              & \multicolumn{1}{c}{}                                 & \multicolumn{1}{c}{}                                  & \multicolumn{1}{c}{}                              & \multicolumn{1}{c}{\# (\%)}           \\ \hline
ActiveMQ                          & 0.0022                                            & 10.95                                                & 125.10                                                & 1,440.44                                          & 17 ( 7.9\%)      \\
Ambari                            & 0.0005                                            & 0.29                                                 & 14.83                                                 & 1,258.75                                          & 1 ( 0.3\%)       \\
Arrow                             & 0.0007                                            & 0.85                                                 & 46.77                                                 & 883.01                                            & 10 ( 3.4\%)      \\
Axis2                             & 0.0007                                            & 15.07                                                & 72.34                                                 & 2,280.70                                          & 12 ( 2.8\%)      \\
Camel                             & 0.0004                                            & 0.50                                                 & 19.66                                                 & 878.04                                            & 11 ( 1.4\%)      \\
CloudStack                        & 0.0013                                            & 5.93                                                 & 30.07                                                 & 1,028.01                                          & 3 ( 0.3\%)       \\
Cordova                           & 0.0010                                            & 5.13                                                 & 54.25                                                 & 977.14                                            & 33 ( 4.1\%)      \\
Drill                             & 0.0007                                            & 12.92                                                & 49.38                                                 & 998.90                                            & 12 ( 2.1\%)      \\
Flink                             & 0.0002                                            & 8.20                                                 & 180.29                                                & 2,452.13                                          & 280 (13.7\%)    \\
Geode                             & 0.0001                                            & 5.20                                                 & 170.92                                                & 1,157.41                                          & 29 (19.3\%)     \\
Groovy                            & 0.0004                                            & 4.06                                                 & 145.06                                                & 4,024.97                                          & 34 ( 9.7\%)      \\
Guacamole                         & 0.0005                                            & 0.16                                                 & 16.49                                                 & 334.85                                            & 0 ( 0.0\%)       \\
Hadoop                            & 0.0006                                            & 2.92                                                 & 65.17                                                 & 2,163.18                                          & 114 ( 5.1\%)     \\
HBase                             & 0.0004                                            & 1.65                                                 & 49.60                                                 & 2,774.03                                          & 41 ( 3.6\%)      \\
Hive                              & 0.0020                                            & 4.76                                                 & 61.10                                                 & 1,349.05                                          & 24 ( 4.7\%)      \\
Hudi                              & 0.0010                                            & 6.74                                                 & 64.35                                                 & 1,007.95                                          & 13 ( 7.1\%)      \\
Ignite                            & 0.0001                                            & 4.95                                                 & 94.41                                                 & 1,294.59                                          & 54 ( 9.3\%)      \\
Impala                            & 0.0001                                            & 13.07                                                & 113.77                                                & 2,611.83                                          & 145 (10.1\%)    \\
Jackrabbit-Oak                    & 0.0040                                            & 3.27                                                 & 63.69                                                 & 1,533.01                                          & 7 ( 4.2\%)       \\
Kafka                             & 0.0003                                            & 1.75                                                 & 64.45                                                 & 2,372.47                                          & 31 ( 5.0\%)      \\
Lucene                            & 0.0015                                            & 2.84                                                 & 211.36                                                & 2,028.66                                          & 42 (20.9\%)     \\
Mesos                             & 0.0008                                            & 5.07                                                 & 90.36                                                 & 1,189.66                                          & 35 ( 8.2\%)      \\
NetBeans                          & 0.0013                                            & 7.24                                                 & 61.79                                                 & 1,184.81                                          & 14 ( 4.5\%)      \\
NiFi                              & 0.0002                                            & 0.85                                                 & 33.21                                                 & 978.16                                            & 5 ( 2.3\%)       \\
OFBiz                             & 0.0004                                            & 2.46                                                 & 108.93                                                & 2,144.06                                          & 22 (10.2\%)     \\
Ozone                             & 0.0014                                            & 3.44                                                 & 55.37                                                 & 1,246.04                                          & 8 ( 4.1\%)       \\
Qpid                              & 0.0003                                            & 4.07                                                 & 45.29                                                 & 843.26                                            & 8 ( 3.7\%)       \\
Solr                              & 0.0005                                            & 4.64                                                 & 81.78                                                 & 2,122.48                                          & 24 ( 6.4\%)      \\
Spark                             & 0.0001                                            & 0.92                                                 & 27.31                                                 & 1,453.73                                          & 28 ( 1.6\%)      \\
Thrift                            & 0.0035                                            & 7.73                                                 & 171.36                                                & 2,642.75                                          & 25 (13.7\%)     \\
Traffic Server                    & 0.0006                                            & 3.15                                                 & 56.11                                                 & 944.63                                            & 6 ( 3.6\%)       \\
Wicket                            & 0.0007                                            & 0.30                                                 & 20.31                                                 & 1,652.07                                          & 1 ( 0.5\%)       \\ \hline
\end{tabular}
}
\end{table}

\begin{table}[htb]
\centering
\caption{Distribution of change phases of bug priority changes (RQ2).}
\label{table:changephase}
\scalebox{0.62}{
\begin{tabular}{lrrrrr}
\hline
\multirow{2}{*}{\textbf{Project}} & \multicolumn{1}{c}{\multirow{2}{*}{\textbf{Total}}} & \multicolumn{1}{c}{\textbf{BEFORE}}                             & \multicolumn{1}{c}{\textbf{PROGRESS}}                        & \multicolumn{1}{c}{REOPEN}         & \multicolumn{1}{c}{AFTER}         \\ \cline{3-6} 
                                  & \multicolumn{1}{c}{}                                & \multicolumn{1}{c}{\# (\%)}                                       & \multicolumn{1}{c}{\# (\%)}                                    & \multicolumn{1}{c}{\# (\%)}          & \multicolumn{1}{c}{\# (\%)}         \\ \hline
ActiveMQ                          & 231                                                 & 101 (43.7\%)                              & \textbf{114 (49.4\%)} & 13 ( 5.6\%)   & 3 ( 1.3\%)   \\
Ambari                            & 317                                                 & 97 (30.6\%)                               & \textbf{163 (51.4\%)} & 13 ( 4.1\%)   & 44 (13.9\%) \\
Arrow                             & 316                                                 & \textbf{227 (71.8\%)}    & 78 (24.7\%)                            & 3 ( 0.9\%)    & 8 ( 2.5\%)   \\
Axis2                             & 503                                                 & 189 (37.6\%)                              & \textbf{277 (55.1\%)} & 33 ( 6.6\%)   & 4 ( 0.8\%)   \\
Camel                             & 867                                                 & \textbf{433 (49.9\%)}    & 374 (43.1\%)                           & 13 ( 1.5\%)   & 47 ( 5.4\%)  \\
CloudStack                        & 1,027                                               & 417 (40.6\%)                              & \textbf{463 (45.1\%)} & 111 (10.8\%) & 36 ( 3.5\%)  \\
Cordova                           & 916                                                 & \textbf{488 (53.3\%)}    & 388 (42.4\%)                           & 28 ( 3.1\%)   & 12 ( 1.3\%)  \\
Drill                             & 644                                                 & \textbf{332 (51.6\%)}    & 294 (45.7\%)                           & 17 ( 2.6\%)   & 1 ( 0.2\%)   \\
Flink                             & 2,629                                               & \textbf{1,544 (58.7\%)}  & 767 (29.2\%)                           & 130 ( 4.9\%)  & 188 ( 7.2\%) \\
Geode                             & 158                                                 & \textbf{71 (44.9\%)}     & 66 (41.8\%)                            & 18 (11.4\%)  & 3 ( 1.9\%)   \\
Groovy                            & 381                                                 & \textbf{181 (47.5\%)}    & 176 (46.2\%)                           & 19 ( 5.0\%)   & 5 ( 1.3\%)   \\
Guacamole                         & 226                                                 & \textbf{171 (75.7\%)}    & 49 (21.7\%)                            & 6 ( 2.7\%)    & 0 ( 0.0\%)   \\
Hadoop                            & 2,507                                               & \textbf{1,262 (50.3\%)}  & 1,030 (41.1\%)                         & 70 ( 2.8\%)   & 145 ( 5.8\%) \\
HBase                             & 1,272                                               & \textbf{615 (48.3\%)}    & 535 (42.1\%)                           & 50 ( 3.9\%)   & 72 ( 5.7\%)  \\
Hive                              & 549                                                 & \textbf{264 (48.1\%)}    & 254 (46.3\%)                           & 15 ( 2.7\%)   & 16 ( 2.9\%)  \\
Hudi                              & 258                                                 & \textbf{134 (51.9\%)}    & 114 (44.2\%)                           & 6 ( 2.3\%)    & 4 ( 1.6\%)   \\
Ignite                            & 674                                                 & \textbf{349 (51.8\%)}    & 301 (44.7\%)                           & 11 ( 1.6\%)   & 13 ( 1.9\%)  \\
Impala                            & 1,792                                               & \textbf{872 (48.7\%)}    & 826 (46.1\%)                           & 72 ( 4.0\%)   & 22 ( 1.2\%)  \\
Jackrabbit-Oak                    & 181                                                 & \textbf{82 (45.3\%)}     & 68 (37.6\%)                            & 7 ( 3.9\%)    & 24 (13.3\%) \\
Kafka                             & 701                                                 & \textbf{447 (63.8\%)}    & 199 (28.4\%)                           & 31 ( 4.4\%)   & 24 ( 3.4\%)  \\
Lucene                            & 213                                                 & 86 (40.4\%)                               & \textbf{111 (52.1\%)} & 10 ( 4.7\%)   & 6 ( 2.8\%)   \\
Mesos                             & 510                                                 & \textbf{270 (52.9\%)}    &215 (42.2\%)                               & 1 ( 0.2\%)    & 24 ( 4.7\%)  \\
NetBeans                          & 377                                                 & 164 (43.5\%)                              & \textbf{204 (54.1\%)}     & 7 ( 1.9\%)    & 2 ( 0.5\%)   \\
NiFi                              & 239                                                 & \textbf{145 (60.7\%)}    & 74 (31.0\%)                               & 11 ( 4.6\%)   & 9 ( 3.8\%)   \\
OFBiz                             & 238                                                 & 107 (45.0\%)                              & \textbf{108 (45.4\%)}     & 13 ( 5.5\%)   & 10 ( 4.2\%)  \\
Ozone                             & 209                                                 & \textbf{141 (67.5\%)}    & 59 (28.2\%)                               & 1 ( 0.5\%)    & 8 ( 3.8\%)   \\
Qpid                              & 230                                                 & \textbf{104 (45.2\%)}    & 95 (41.3\%)                               & 8 ( 3.5\%)    & 23 (10.0\%) \\
Solr                              & 413                                                 & \textbf{217 (52.5\%)}    & 165 (40.0\%)                               & 18 ( 4.4\%)   & 13 ( 3.1\%)  \\
Spark                             & 2,028                                               & \textbf{965 (47.6\%)}    & 819 (40.4\%)                               & 62 ( 3.1\%)   & 182 ( 9.0\%) \\
Thrift                            & 202                                                 & 85 (42.1\%)                               & \textbf{82 (45.5\%)}     & 5 ( 2.5\%)    & 20 ( 9.9\%)  \\
Traffic Server                    & 179                                                 & \textbf{94 (52.5\%)}     & 74 (41.3\%)                               & 4 ( 2.2\%)    & 7 ( 3.9\%)   \\
Wicket                            & 228                                                 & \textbf{104 (45.6\%)}    & 95 (41.7\%)                               & 12 ( 5.3\%)   & 17 ( 7.5\%)  \\
All projects                      & 21,215                                              & \textbf{10,758 (50.7\%)} & 8,647 (40.8\%)                               & 818 ( 3.9\%)  & 992 ( 4.7\%)  \\ \hline
\end{tabular}
}
\end{table}

\subsection{Time Characteristics of Bug Priority Changes (RQ2)}
\subsubsection{Time Intervals of Bug Priority Changes}
We first investigated the time intervals between when the bugs were reported and when their priorities were changed for the first time, and the result is shown in Table \ref{table:ChangeTimeInterval}. In each project, the minimum time interval of bugs ranges from 0.0001 to 0.0040 days, i.e., 9 to 346 seconds, and the average time interval of bugs ranges from 14.83 to 211.36 days, while the median time interval ranges from 0.16 to 15.07 days, which is way shorter than the average. In addition, the median time interval of priority changes is merely a few days for all projects except for projects \textit{ActiveMQ}, \textit{Axis2}, \textit{Drill} and \textit{Impala}.

As shown in Table \ref{table:ChangeTimeInterval}, the time intervals of priority changes of some bugs are more than 365 days. For most (19 out of 32) of projects, the time intervals of less than 5.0\% of the bugs are more than 365 days. For 6 of the projects, the time intervals of more than 10.0\% of the bugs are more than 365 days.

\subsubsection{Change Phases of Bug Priority}
The distribution of change phases of bug priority changes is shown in Table \ref{table:changephase}. For 75\% of (24 out of 32) projects, bug priority changes that happened in change phase BEFORE (i.e., before the bug was handled) account for the most. For the remaining projects (8 out of 32), bug priority changes that happened in change phase PROGRESS (i.e., during the period when the bug was being handled) account for the most. In addition, for all the projects, only a very small proportion of bug priority changes were made in change phases REOPEN (i.e., after the bug was reopened) and AFTER (i.e., after the bug's status was set as Closed or Resolved) respectively.

\begin{table}[]
\centering
\caption{Distribution of bugs over the number of priority changes (RQ3).}
\label{table:changetimes}
\scalebox{0.65}{
\begin{tabular}{lrrrrr}
\hline
\multirow{2}{*}{\textbf{Project}} & \multicolumn{1}{c}{\multirow{2}{*}{\textbf{Total}}} & \multicolumn{1}{c}{\textbf{C1}}                                 & \multicolumn{1}{c}{C2}              & \multicolumn{1}{c}{C3}            & \multicolumn{1}{c}{C3+}          \\ \cline{3-6} 
                                  & \multicolumn{1}{c}{}                                & \multicolumn{1}{c}{\# (\%)}                                     & \multicolumn{1}{c}{\# (\%)}         & \multicolumn{1}{c}{\# (\%)}       & \multicolumn{1}{c}{\# (\%)}      \\ \hline
ActiveMQ                          & 216                                                 & \textbf{205 (94.9\%)}    & 9 ( 4.2\%)     & 1 ( 0.5\%)   & 1 ( 0.5\%)  \\
Ambari                            & 307                                                 & \textbf{299 (97.4\%)}    & 6 ( 2.0\%)     & 2 ( 0.7\%)   & 0 ( 0.0\%)  \\
Arrow                             & 290                                                 & \textbf{268 (92.4\%)}    & 19 ( 6.6\%)    & 2 ( 0.7\%)   & 1 ( 0.3\%)  \\
Axis2                             & 425                                                 & \textbf{360 (84.7\%)}    & 56 (13.2\%)   & 6 ( 1.4\%)   & 3 ( 0.7\%)  \\
Camel                             & 807                                                 & \textbf{767 (95.0\%)}    & 27 ( 3.3\%)    & 11 ( 1.4\%)  & 2 ( 0.2\%)  \\
CloudStack                        & 870                                                 & \textbf{745 (85.6\%)}    & 99 (11.4\%)   & 22 ( 2.5\%)  & 4 ( 0.5\%)  \\
Cordova                           & 813                                                 & \textbf{730 (89.8\%)}    & 69 ( 8.5\%)    & 9 ( 1.1\%)   & 5 ( 0.6\%)  \\
Drill                             & 579                                                 & \textbf{515 (88.9\%)}    & 63 (10.9\%)   & 1 ( 0.2\%)   & 0 ( 0.0\%)  \\
Flink                             & 2,049                                               & \textbf{1,630 (79.6\%)}  & 299 (14.6\%)  & 96 ( 4.7\%)  & 24 ( 1.2\%) \\
Geode                             & 150                                                 & \textbf{142 (94.7\%)}    & 8 ( 5.3\%)     & 0 ( 0.0\%)   & 0 ( 0.0\%)  \\
Groovy                            & 352                                                 & \textbf{325 (92.3\%)}    & 25 ( 7.1\%)    & 2 ( 0.6\%)   & 0 ( 0.0\%)  \\
Guacamole                         & 209                                                 & \textbf{197 (94.3\%)}    & 8 ( 3.8\%)     & 3 ( 1.4\%)   & 1 ( 0.5\%)  \\
Hadoop                            & 2,244                                               & \textbf{2,015 (89.8\%)}  & 202 ( 9.0\%)   & 21 ( 0.9\%)  & 6 ( 0.3\%)  \\
HBase                             & 1,142                                               & \textbf{1,028 (90.0\%)}  & 101 ( 8.8\%)   & 10 ( 0.9\%)  & 3 ( 0.3\%)  \\
Hive                              & 509                                                 & \textbf{472 (92.7\%)}    & 34 ( 6.7\%)    & 3 ( 0.6\%)   & 0 ( 0.0\%)  \\
Hudi                              & 184                                                 & \textbf{130 (70.7\%)}    & 39 (21.2\%)   & 13 ( 7.1\%)  & 2 ( 1.1\%)  \\
Ignite                            & 582                                                 & \textbf{505 (86.8\%)}    & 65 (11.2\%)   & 9 ( 1.5\%)   & 3 ( 0.5\%)  \\
Impala                            & 1,431                                               & \textbf{1,168 (81.6\%)}  & 196 (13.7\%)  & 45 ( 3.1\%)  & 22 ( 1.5\%) \\
Jackrabbit-Oak                    & 168                                                 & \textbf{156 (92.9\%)}    & 11 ( 6.5\%)    & 1 ( 0.6\%)   & 0 ( 0.0\%)  \\
Kafka                             & 617                                                 & \textbf{545 (88.3\%)}    & 60 ( 9.7\%)    & 12 ( 1.9\%)  & 0 ( 0.0\%)  \\
Lucene                            & 201                                                 & \textbf{191 (95.0\%)}    & 8 ( 4.0\%)     & 2 ( 1.0\%)   & 0 ( 0.0\%)  \\
Mesos                             & 427                                                 & \textbf{364 (85.2\%)}    & 46 (10.8\%)   & 15 ( 3.5\%)  & 2 ( 0.5\%)  \\
NetBeans                          & 314                                                 & \textbf{259 (82.5\%)}    & 48 (15.3\%)   & 6 ( 1.9\%)   & 1 ( 0.3\%)  \\
NiFi                              & 221                                                 & \textbf{205 (92.8\%)}    & 15 ( 6.8\%)    & 0 ( 0.0\%)   & 1 ( 0.5\%)  \\
OFBiz                             & 215                                                 & \textbf{196 (91.2\%)}    & 15 ( 7.0\%)    & 4 ( 1.9\%)   & 0 ( 0.0\%)  \\
Ozone                             & 194                                                 & \textbf{180 (92.8\%)}    & 13 ( 6.7\%)    & 1 ( 0.5\%)   & 0 ( 0.0\%)  \\
Qpid                              & 214                                                 & \textbf{199 (93.0\%)}    & 14 ( 6.5\%)    & 1 ( 0.5\%)   & 0 ( 0.0\%)  \\
Solr                              & 377                                                 & \textbf{347 (92.0\%)}    & 25 ( 6.6\%)    & 4 ( 1.1\%)   & 1 ( 0.3\%)  \\
Spark                             & 1,773                                               & \textbf{1,556 (87.8\%)}  & 185 (10.4\%)  & 26 ( 1.5\%)  & 6 ( 0.3\%)  \\
Thrift                            & 183                                                 & \textbf{167 (91.3\%)}    & 14 ( 7.7\%)    & 1 ( 0.5\%)   & 1 ( 0.5\%)  \\
Traffic Server                    & 165                                                 & \textbf{152 (92.1\%)}    & 12 ( 7.3\%)    & 1 ( 0.6\%)   & 0 ( 0.0\%)  \\
Wicket                            & 212                                                 & \textbf{198 (93.4\%)}    & 12 ( 5.7\%)    & 2 ( 0.9\%)   & 0 ( 0.0\%)  \\
All projects                      & 18,440                                              & \textbf{16,216 (87.9\%)} & 1,803 ( 9.8\%) & 332 ( 1.8\%) & 89 ( 0.5\%) \\ \hline
\end{tabular}
}
\end{table}

\subsection{Patterns of Bug Priority Changes (RQ3)}
In this subsection, we first calculate the number of priority changes of bugs in each project, and then investigate the priority change patterns of bugs in each project. Finally, we show the trend of priority changes and the distribution of the average change range of each bug, and explore the relationship between priority change and priority itself.

\subsubsection{Number of Priority Changes of Bugs}
Table \ref{table:changetimes} shows the distribution of bugs over the number of priority changes for each project. For all projects, most (at least 70.7\%) bugs underwent only one priority change, and a small proportion (no more than 21.2\%) of bugs underwent two priority changes. Additionally, a small proportion of bugs underwent three or more priority changes in 31 out of the 32 projects, and the other project (i.e., \textit{Geode}) does not have any bugs with three or more priority changes.
When taking all projects as a whole, 87.9\% of the total bugs underwent only one priority change, 9.8\% of the total bugs underwent two priority changes, 1.8\% of the total bugs underwent three priority changes, and only 0.5\% of the total bugs underwent more than three priority changes.

\begin{table*}[]
\centering
\caption{Distribution of bugs over priority change patterns (RQ3).}
\label{table:patterndistribution}
\scalebox{0.5}{
} & \cellcolor[HTML]{C0C0C0}\textbf{32}                                                        & \cellcolor[HTML]{C0C0C0}\textbf{32}                                                        & \cellcolor[HTML]{C0C0C0}\textbf{27}                                                         & \cellcolor[HTML]{C0C0C0}\textbf{23}                                                         & \cellcolor[HTML]{C0C0C0}\textbf{14}                                                         & \cellcolor[HTML]{C0C0C0}\textbf{24}                                                         & \cellcolor[HTML]{C0C0C0}\textbf{27}                                                         & \cellcolor[HTML]{C0C0C0}\textbf{13}                                                         & \cellcolor[HTML]{C0C0C0}\textbf{32}                                                         & \cellcolor[HTML]{C0C0C0}\textbf{30}                                                         & \cellcolor[HTML]{C0C0C0}\textbf{21}                                                        & \cellcolor[HTML]{C0C0C0}\textbf{19}                                                         & \cellcolor[HTML]{C0C0C0}\textbf{12}                                                         & \cellcolor[HTML]{C0C0C0}\textbf{1}                                                           & \cellcolor[HTML]{C0C0C0}\textbf{3}                                                           & \cellcolor[HTML]{C0C0C0}\textbf{17}                                                        & \cellcolor[HTML]{C0C0C0}\textbf{14}                                                         & \cellcolor[HTML]{C0C0C0}\textbf{12}                                                         & \cellcolor[HTML]{C0C0C0}\textbf{1}                                                           & \cellcolor[HTML]{C0C0C0}\textbf{1}                                                           & \cellcolor[HTML]{C0C0C0}\textbf{15}                                                         & \cellcolor[HTML]{C0C0C0}\textbf{6}                                                          & \cellcolor[HTML]{C0C0C0}\textbf{8}                                                           & \cellcolor[HTML]{C0C0C0}\textbf{5}                                                           & \cellcolor[HTML]{FFFFFF}                                                                                            \\ \cline{1-25}
\end{tabular}
}
\end{table*}

\subsubsection{Bug Priority Change Patterns}

We consider all possible cases of priority changes of a bug and propose 24 priority change patterns (shown in Figure \ref{fig:Patterns12} and Figure \ref{fig:Patterns3Plus}) of a bug. 
The distribution of the number of bugs of each project over different priority change patterns is shown in Table \ref{table:patterndistribution}, in which each cell lists the number and percentage of bugs with the priority change pattern in the corresponding row. The cell in the far right column lists the number of patterns for each project, and the last row lists the number of projects for each pattern.

As shown in Table \ref{table:patterndistribution}, the distribution of bugs over priority change patterns are quite different from project to project. Specifically, from the perspective of the coverage of the bug priority change patterns (i.e., the rightmost column in Table \ref{table:patterndistribution}), none of the projects covers all the bug priority change patterns, and the number of bug priority change patterns for all the projects falls in the range between 6 (projects \textit{Geode} and \textit{Traffic Server}) and 22 (projects \textit{Flink} and \textit{Impala}).


As shown in the second to last row of Table \ref{table:patterndistribution}, over the 18,440 bugs with priority changes in total of all the projects, there are 8,672 (47.5\%) and 7,544 (40.9\%) bugs following priority change patterns P1I\_I and P1D\_D, respectively, which are way more than bugs with other patterns.


The last row in Table \ref{table:patterndistribution} shows the number of projects covering each bug priority change patterns. In the last row, P1I\_I, P1D\_D, and P2R\_ID are covered by all projects, 8 patterns (i.e., P2I\_II, P2I\_ID, P2D\_DD, P2D\_ID, P2R\_DI, PmI\_I, PmI\_ID, and PmD\_D) are covered by more than a half of the projects, 3 patterns (i.e., PmI\_IDI, PmD\_IDI, and PmD\_DID) are covered by only one project.



\subsubsection{Bug Priority Change Trend and Range}

\begin{figure}
    \centering{\includegraphics[width=3.3in]{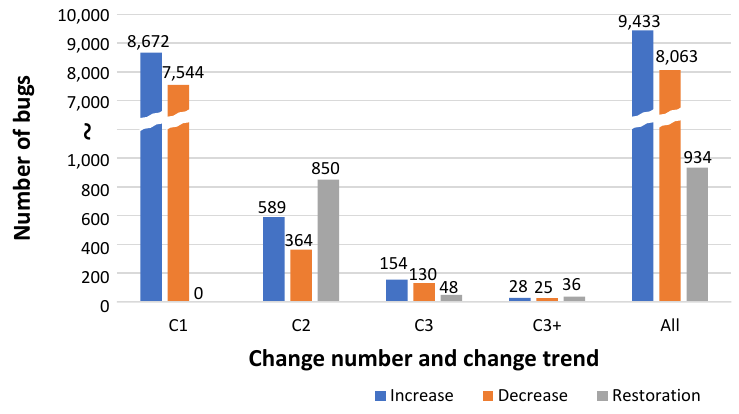}}
    \caption{Distribution of the number of bugs over priority change trend and the number of changes (RQ3).}
    \label{fig:PatternTypeBug}
\end{figure}

Figure \ref{fig:PatternTypeBug} shows the trend of bugs with different numbers of priority changes. Priorities of 9,443 out of 18,440 (51.2\%) bugs increase finally; priorities of 8,063 out of 18,440 (43.7\%) bugs decrease finally; and priorities of 934 out of 18,440 (5.1\%) bugs restore finally. In addition, 850 of 934 (91.0\%) bugs that priority restore finally come from bugs that undergo two priority changes.

\begin{figure}
    \centering{\includegraphics[width=3.3in]{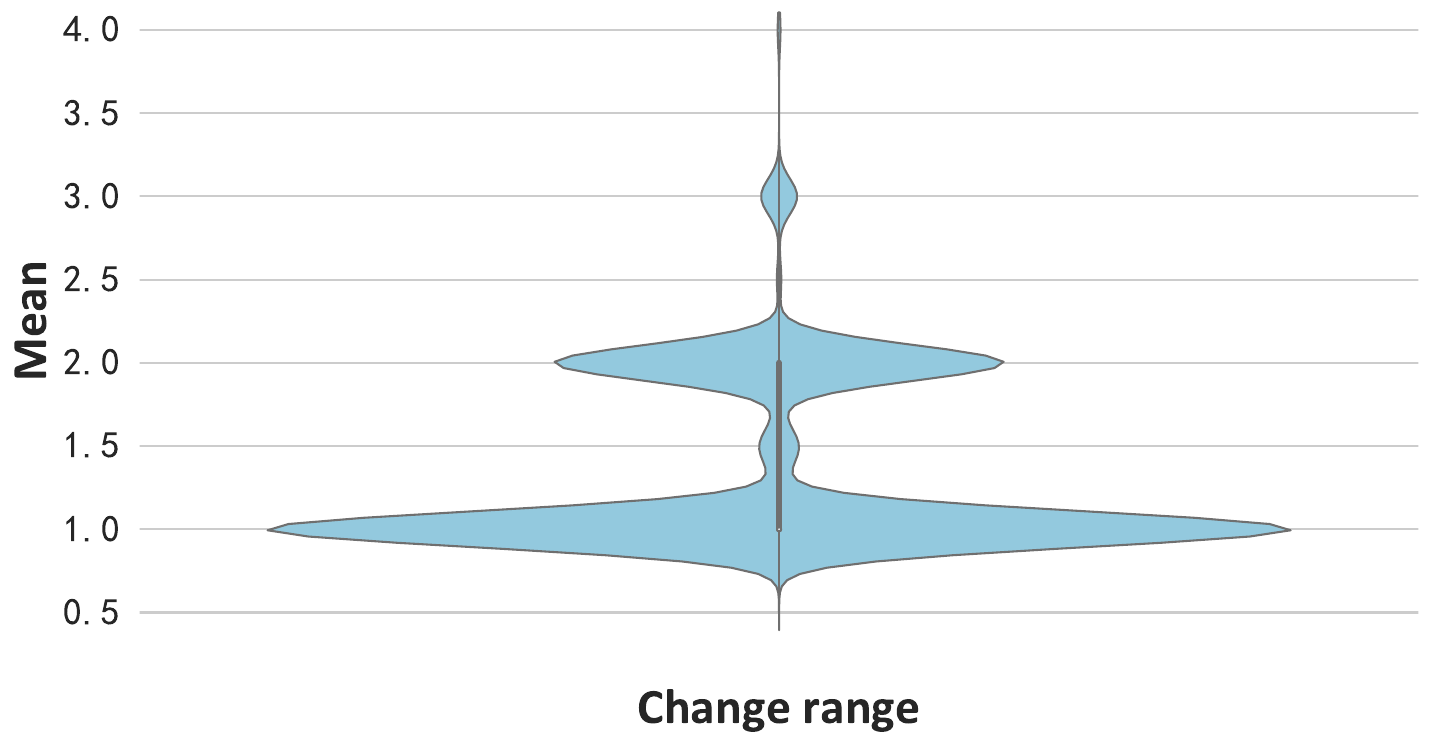}}
    \caption{Distribution of average bug change range (RQ3).}
    \label{fig:ChangeRange}
\end{figure}

Figure \ref{fig:ChangeRange} shows the distribution of average priority change ranges of each bug for all projects (i.e., 18,440 bugs). For most of the bugs, their average priority change ranges are 1.0; the average priority change ranges of a minority of bugs are 2.0; and bugs with an average priority change ranges of 1.5 or 3.0 are rare.


\begin{table*}[]
\centering
\caption{Distribution of priority change probability of different priorities (RQ3).}
\label{table:priorityiteself}
\scalebox{0.78}{
\begin{tabular}{lrrrrrrrrrrrrrrr}
\hline
                                   & \multicolumn{3}{c}{\textbf{Blocker}}                                                                    & \multicolumn{3}{c}{\textbf{Critical}}                                                                   & \multicolumn{3}{c}{\textbf{Major}}                                                                      & \multicolumn{3}{c}{\textbf{Minor}}                                                                      & \multicolumn{3}{c}{\textbf{Trivial}}                                                                    \\ \cline{2-16} 
\multirow{-2}{*}{\textbf{Project}} & \multicolumn{1}{c}{\textbf{\#OR}} & \multicolumn{1}{c}{\textbf{\#PC}} & \multicolumn{1}{c}{\textbf{\%PC}} & \multicolumn{1}{c}{\textbf{\#OR}} & \multicolumn{1}{c}{\textbf{\#PC}} & \multicolumn{1}{c}{\textbf{\%PC}} & \multicolumn{1}{c}{\textbf{\#OR}} & \multicolumn{1}{c}{\textbf{\#PC}} & \multicolumn{1}{c}{\textbf{\%PC}} & \multicolumn{1}{c}{\textbf{\#OR}} & \multicolumn{1}{c}{\textbf{\#PC}} & \multicolumn{1}{c}{\textbf{\%PC}} & \multicolumn{1}{c}{\textbf{\#OR}} & \multicolumn{1}{c}{\textbf{\#PC}} & \multicolumn{1}{c}{\textbf{\%PC}} \\ \hline
ActiveMQ                           & 191                               & 43                                & \textbf{22.5\%}                   & 451                               & 53                                & 11.8\%                            & 3,835                              & 116                               & 3.0\%                             & 718                               & 18                                & 2.5\%                             & 94                                & 1                                 & 1.1\%                             \\
\hline
\end{tabular}%
}
\end{table*}

\subsubsection{Relationship Between Bug Priority Change and Priority Itself}

The probability of priority change of different priorities is listed in Table \ref{table:priorityiteself}. In this table, \#OR denotes the number of original priorities\footnote{Note that the statistical unit here is a change record but not a bug, because a bug may have multiple priority changes.}, \#PC denotes the number of priority changes. For 20 of the 32 projects, Blocker has the highest probability of priority change, and Critical has the highest probability of priority change in 10 of the remaining 12 projects. For all projects, as shown in Table \ref{table:proportion}, the priority change probability of Blocker is 16.1\%, which is the highest of all priorities and much higher than 8.3\% (i.e., the percentage of bugs with priority changes over all bugs of all the 32 projects). In addition, the reason why the priority change probability of Trivial in project \textit{Flink} is 100\% is that all the bug priority changes were made in a batch priority modification (i.e., the priorities of all bugs with priority changes were modified by someone at a time point). After removing the changes in \textit{Flink}'s Trivial, we conducted a Spearman correlation coefficient analysis on the priority and its probability of change. The \textit{p-value} is less than 0.05 and the correlation coefficient is 0.99, indicating a significant strong correlation between bug priority and the probability of priority change, i.e., the higher the bug priority, the greater the probability of bug priority change.

Table \ref{table:priorityiteselfmatrix} lists the number of pairwise changes between different priorities of all projects. The left is the original priority, and the top is the changed priority. As we can see, most priorities are changed to the adjacent priorities; Major has the highest original priority, accounting for 60.1\% (12,746 out of 21,215); 
except for Trivial, the numbers of changed priorities of all projects distribute over Blocker (5,398), Critical (5,612), Major (4,329), and Minor (5,318) in a balanced way in the sense that there is no big difference between the numbers. 

\begin{table}[]
\centering
\caption{Distribution of pairwise changes with different priorities (RQ3).}
\label{table:priorityiteselfmatrix}
\scalebox{0.82}{
\begin{tabular}{lrrrrrr}
\hline
                  & \multicolumn{1}{c}{\textbf{Blocker}} & \multicolumn{1}{c}{\textbf{Critical}} & \multicolumn{1}{c}{\textbf{Major}} & \multicolumn{1}{c}{\textbf{Minor}} & \multicolumn{1}{c}{\textbf{Trivial}} & \multicolumn{1}{c}{\textbf{All}} \\
\rowcolor[HTML]{EFEFEF} 
\textbf{Blocker}  & 0                                    & 1,019                                  & 1,480                               & 298                                & 29                                   & 2,826                             \\
\textbf{Critical} & 1,419                                 & 0                                     & 1,789                               & 370                                & 19                                   & 3,597                             \\
\rowcolor[HTML]{EFEFEF} 
\textbf{Major}    & 3,755                                 & 4,266                                  & 0                                  & 4,383                               & 342                                  & 12,746                            \\
\textbf{Minor}    & 209                                  & 312                                   & 988                               & 0                                  & 168                                  & 1,677                             \\
\rowcolor[HTML]{EFEFEF} 
\textbf{Trivial}  & 15                                   & 15                                    & 72                                 & 267                                & 0                                    & 369                              \\
\textbf{All}      & 5,398                                 & 5,612                                  & 4,329                               & 5,318                               & 558                                  & 21,215                            \\ \hline
\end{tabular}%
}
\end{table}

\subsection{Complexity of Bugs (RQ4)}
In this subsection, we calculate change complexity of bug-fixing commits and communication complexity of bugs with/without priority changes, in order to explore whether there is a significant difference on complexity between bugs with priority changes and bugs without priority changes. The bug-fixing commits of bugs and comments of bugs can respectively reflect the complexity of bugs from the perspective of code change and bug participants.

\subsubsection{Change Complexity of the Bug-fixing Commits for Bugs}

Table \ref{table:changecomplexity} shows the results of the Mann-Whitney U tests on the change complexity of bug-fixing commits for bugs with priority changes and bugs without priority changes. In this table, AveC denotes the average value of the corresponding change complexity metric of bug-fixing commits for bugs with priority changes and AveN denotes that of bugs without priority changes. 
\textit{P-value} followed by $\star$ denotes that AveC is significantly less than AveN. As shown in Table \ref{table:changecomplexity}, for change complexity metrics LOCM, NOFM, NOPM, and Entropy, the AveC of 29, 21, 24, and 25 out of 32 projects respectively are significantly greater than the corresponding AveN.


\begin{table*}[]
\centering
\caption{Average change complexity of bug-fixing commits for bugs in the selected projects (RQ4).}
\label{table:changecomplexity}
\scalebox{0.6}{
\begin{tabular}{lrrrrrrrrrrrrrrrr}
\hline
                                   & \multicolumn{4}{c}{\textbf{LOCM}}                                                                                     & \multicolumn{4}{c}{\textbf{NOFM}}                                                                                     & \multicolumn{4}{c}{\textbf{NOPM}}                                                                                     & \multicolumn{4}{c}{\textbf{Entropy}}                                                                                  \\ \cline{2-17} 
\multirow{-2}{*}{\textbf{Project}} & \multicolumn{1}{c}{\textbf{AveC}} & \multicolumn{1}{c}{\textbf{AveN}} & \multicolumn{1}{c}{\textit{\textbf{p-value}}} & \multicolumn{1}{c}{\textit{\textbf{\begin{tabular}[c]{@{}c@{}}Effect\\  Size\end{tabular}}}} & \multicolumn{1}{c}{\textbf{AveC}} & \multicolumn{1}{c}{\textbf{AveN}} & \multicolumn{1}{c}{\textit{\textbf{p-value}}} & \multicolumn{1}{c}{\textit{\textbf{\begin{tabular}[c]{@{}c@{}}Effect\\  Size\end{tabular}}}} & \multicolumn{1}{c}{\textbf{AveC}} & \multicolumn{1}{c}{\textbf{AveN}} & \multicolumn{1}{c}{\textit{\textbf{p-value}}} & \multicolumn{1}{c}{\textit{\textbf{\begin{tabular}[c]{@{}c@{}}Effect\\  Size\end{tabular}}}} & \multicolumn{1}{c}{\textbf{AveC}} & \multicolumn{1}{c}{\textbf{AveN}} & \multicolumn{1}{c}{\textit{\textbf{p-value}}} & \multicolumn{1}{c}{\textit{\textbf{\begin{tabular}[c]{@{}c@{}}Effect\\  Size\end{tabular}}}} \\ \hline
ActiveMQ                           & 189                               & 162                               & \textless{}0.001     & 0.05                         & 3.21                              & 3.57                              & \textless{}0.001$\star$   & 0.08   & 2.25                              & 2.42                              & \textless{}0.001$\star$   & 0.07   & 0.60                              & 0.62                              & \textless{}0.001$\star$  & 0.17    \\
Ambari                             & 142                               & 143                               & \textless{}0.001$\star$  & 0.02    & 4.13                              & 4.09                              & \textless{}0.001        & 0.02                      & 2.87                              & 2.86         & \textless{}0.001                     & \textless{}0.01                              & 0.63                              & 0.59        &  \textless{}0.001                      & 0.01                              \\
Arrow                              & 189                               & 110                               & \textless{}0.001    & 0.22                          & 3.65                              & 3.52                              & \textless{}0.001              & 0.07                & 1.87                              & 1.84                              & \textless{}0.001          & 0.02                    & 0.68                              & 0.61                              & \textless{}0.001         & 0.09                     \\
Axis2                              & 212                               & 148                               & \textless{}0.001    & 0.19                          & 3.44                              & 3.48                              & \textless{}0.001$\star$  & 0.01    & 2.50                              & 2.32                              & \textless{}0.001           & 0.08                   & 0.53                              & 0.47                              & \textless{}0.001        & 0.19                      \\
Camel                              & 120                               & 106                               & \textless{}0.001   & 0.07                           & 3.36                              & 3.55                              & \textless{}0.001$\star$   & 0.05   & 2.38                              & 2.39                              & \textless{}0.001$\star$  & 0.01    & 0.70                              & 0.65                              & \textless{}0.001           & 0.11                   \\
CloudStack                         & 99                                & 94                                & \textless{}0.001    & \textless{}0.01                          & 2.95                              & 2.89                              & \textless{}0.001              & 0.01                & 2.18                              & 2.19                              & \textless{}0.001$\star$  & \textless{}0.01    & 0.39                              & 0.35                              & \textless{}0.001           & 0.14                   \\
Cordova                            & 54                                & 114                               & \textless{}0.001$\star$  & 0.07    & 1.95                              & 2.56                              & \textless{}0.001$\star$  & 0.11    & 1.46                              & 1.60                              & \textless{}0.001$\star$  & 0.05    & 0.38                              & 0.40                              & \textless{}0.001$\star$  & \textless{}0.01    \\
Drill                              & 308                               & 201                               & \textless{}0.001    & 0.14                          & 5.87                              & 5.18                              & \textless{}0.001        & 0.08                      & 3.76                              & 3.28                              & \textless{}0.001          & 0.09                    & 0.73                              & 0.66                              & \textless{}0.001           & 0.05                   \\
Flink                              & 213                               & 181                               & \textless{}0.001      & 0.08                        & 5.11                              & 4.65                              & \textless{}0.001          & 0.08                    & 3.32                              & 3.10                              & \textless{}0.001            & 0.07                  & 0.65                              & 0.66                              & \textless{}0.001$\star$   & 0.02   \\
Geode                              & 367                               & 271                               & \textless{}0.001     & 0.16                         & 4.00                              & 4.95                              & \textless{}0.001$\star$   & 0.13   & 2.80                              & 3.08                              & \textless{}0.001$\star$   & 0.09   & 0.62                              & 0.65                              & \textless{}0.001$\star$   & 0.08   \\
Groovy                             & 102                               & 101                               & \textless{}0.001     & 0.02                         & 2.80                              & 2.69                              & \textless{}0.001        & 0.05                      & 2.27                              & 2.21                              & \textless{}0.001            & 0.04                  & 0.61                              & 0.65                              & \textless{}0.001$\star$    & 0.03  \\
Guacamole                          & 139                               & 140                               & \textless{}0.001$\star$  & 0.06    & 3.50                              & 3.00                              & \textless{}0.001        & 0.14                      & 1.74                              & 1.88                              & \textless{}0.001$\star$   & 0.02   & 0.43                              & 0.39                              & \textless{}0.001             & 0.01                 \\
Hadoop                             & 112                               & 94                                & \textless{}0.001      & 0.05                        & 3.83                              & 3.01                              & \textless{}0.001           & 0.13                   & 2.82                              & 2.24                              & \textless{}0.001          & 0.18                    & 0.64                              & 0.55                              & \textless{}0.001            & 0.16                  \\
HBase                              & 178                               & 126                               & \textless{}0.001    & 0.14                          & 4.20                              & 3.44                              & \textless{}0.001        & 0.15                      & 2.59                              & 2.24                              & \textless{}0.001          & 0.16                    & 0.57                              & 0.48                              & \textless{}0.001           & 0.13                   \\
Hive                               & 178                               & 144                               & \textless{}0.001      & 0.07                        & 4.13                              & 3.48                              & \textless{}0.001           & 0.12                   & 2.63                              & 2.30                              & \textless{}0.001            & 0.14                  & 0.55                              & 0.47                              & \textless{}0.001          & 0.21                    \\
Hudi                               & 147                               & 120                               & \textless{}0.001      & 0.14                        & 5.57                              & 4.00                              & \textless{}0.001        & 0.34                      & 4.19                              & 3.11                              & \textless{}0.001          & 0.38                    & 0.74                              & 0.63                              & \textless{}0.001           & 0.36                   \\
Ignite                             & 251                               & 210                               & \textless{}0.001      & 0.19                        & 6.88                              & 5.24                              & \textless{}0.001                    & 0.16          & 4.42                              & 3.46                              & \textless{}0.001              & 0.20                & 0.78                              & 0.64                              & \textless{}0.001           & 0.21                   \\
Impala                             & 105                               & 80                                & \textless{}0.001       & 0.10                       & 3.63                              & 2.94                              & \textless{}0.001          & 0.17                    & 2.16                              & 1.98                              & \textless{}0.001          & 0.14                    & 0.56                              & 0.52                              & \textless{}0.001           & 0.09                   \\
Jackrabbit-Oak                     & 154                               & 138                               & \textless{}0.001      & 0.02                        & 3.71                              & 3.60                              & \textless{}0.001          & 0.02                    & 2.56                              & 2.30                              & \textless{}0.001          & 0.09                    & 0.62                              & 0.60                              & \textless{}0.001          & 0.05                    \\
Kafka                              & 248                               & 161                               & \textless{}0.001      & 0.21                        & 6.41                              & 4.66                              & \textless{}0.001          & 0.22                    & 3.76                              & 2.94                              & \textless{}0.001          & 0.27                    & 0.80                              & 0.69                              & \textless{}0.001           & 0.27                   \\
Lucene                             & 154                               & 146                               & \textless{}0.001      & 0.08                        & 4.03                              & 4.17                              & \textless{}0.001$\star$   & 0.14   & 2.68                              & 2.62                              & \textless{}0.001               & 0.21               & 0.71                              & 0.64                              & \textless{}0.001                & 0.15              \\
Mesos                              & 121                               & 60                                & \textless{}0.001      & 0.36                        & 1.30                              & 1.90                              & \textless{}0.001$\star$   & 0.25   & 1.20                              & 1.34                              & \textless{}0.001$\star$   & 0.20   & 0.17                              & 0.29                              & \textless{}0.001$\star$  & 0.48    \\
NetBeans                           & 217                               & 145                               & \textless{}0.001     & 0.13                         & 3.28                              & 3.43                              & \textless{}0.001$\star$   & 0.10   & 2.39                              & 2.51                              & \textless{}0.001$\star$   & 0.03   & 0.43                              & 0.58                              & \textless{}0.001$\star$    & 0.14  \\
NiFi                               & 209                               & 136                               & \textless{}0.001     & 0.10                         & 4.26                              & 3.47                              & \textless{}0.001                & 0.15              & 2.96                              & 2.45                              & \textless{}0.001          & 0.17                    & 0.61                              & 0.59                              & \textless{}0.001            & 0.20                  \\
OFBiz                              & 38                                & 36                                & \textless{}0.001    & \textless{}0.01                          & 2.43                              & 1.60                              & \textless{}0.001   & 0.18                           & 2.00                              & 1.37                              & \textless{}0.001     & 0.26                         & 0.25                              & 0.20                              & \textless{}0.001         & 0.05                     \\
Ozone                              & 199                               & 149                               & \textless{}0.001    & 0.16                          & 5.67                              & 4.68                              & \textless{}0.001          & 0.17                    & 4.01                              & 3.21                              & \textless{}0.001           & 0.23                   & 0.67                              & 0.56                              & \textless{}0.001             & 0.37                 \\
Qpid                               & 154                               & 147                               & \textless{}0.001     & 0.01                         & 3.98                              & 4.37                              & \textless{}0.001$\star$   & 0.07   & 2.66                              & 2.51                              & \textless{}0.001               & 0.05               & 0.58                              & 0.57                              & \textless{}0.001        & 0.07                      \\
Solr                               & 161                               & 148                               & \textless{}0.001     & 0.13                         & 4.13                              & 4.21                              & \textless{}0.001$\star$   & 0.15   & 2.71                              & 2.62                              & \textless{}0.001          & 0.22                    & 0.73                              & 0.63                              & \textless{}0.001              & 0.14                \\
Spark                              & 94                                & 74                                & \textless{}0.001    & 0.10                          & 3.21                              & 3.16                              & \textless{}0.001        & 0.02                      & 2.47                              & 2.46                              & \textless{}0.001           & 0.02                   & 0.60                              & 0.63                              & \textless{}0.001          & 0.06                    \\
Thrift                             & 125                               & 58                                & \textless{}0.001      & 0.30                        & 3.16                              & 1.95                              & \textless{}0.001         & 0.39                     & 1.96                              & 1.44                              & \textless{}0.001          & 0.47                    & 0.46                              & 0.31                              & \textless{}0.001              & 0.23                \\
Traffic Server                     & 100                               & 99                                & \textless{}0.001    & 0.01                          & 3.23                              & 3.47                              & \textless{}0.001$\star$  & 0.01    & 1.96                              & 1.93                              & \textless{}0.001             & 0.02                 & 0.43                              & 0.41                              & \textless{}0.001            & 0.08                  \\
Wicket                             & 114                               & 90                                & \textless{}0.001      & 0.08                        & 3.21                              & 2.54                              & \textless{}0.001        & 0.21                      & 1.99                              & 1.93                              & \textless{}0.001           & 0.09                   & 0.48                              & 0.46                              & \textless{}0.001           & 0.09                   \\ \hline
\end{tabular}
}
\end{table*}

\begin{table*}[]
\centering
\caption{Average communication complexity for bugs in the selected projects (RQ4)}
\label{table:commentcomplexity}
\scalebox{0.65}{
\begin{tabular}{lrrrrrrrrrrrr}
\hline
                                   & \multicolumn{4}{c}{\textbf{NOC}}                                                                             & \multicolumn{4}{c}{\textbf{TLC}}                                                                            & \multicolumn{4}{c}{\textbf{NOCR}}                                                                       \\ \cline{2-13} 
\multirow{-2}{*}{\textbf{Project}} & \multicolumn{1}{c}{\textbf{AveC}} & \multicolumn{1}{c}{\textbf{AveN}} & \multicolumn{1}{c}{\textit{\textbf{p-value}}} & \multicolumn{1}{c}{\textit{\textbf{\begin{tabular}[c]{@{}c@{}}Effect\\  Size\end{tabular}}}} & \multicolumn{1}{c}{\textbf{AveC}} & \multicolumn{1}{c}{\textbf{AveN}} & \multicolumn{1}{c}{\textit{\textbf{p-value}}} & \multicolumn{1}{c}{\textit{\textbf{\begin{tabular}[c]{@{}c@{}}Effect\\  Size\end{tabular}}}} & \multicolumn{1}{c}{\textbf{AveC}} & \multicolumn{1}{c}{\textbf{AveN}} & \multicolumn{1}{c}{\textit{\textbf{p-value}}} & \multicolumn{1}{c}{\textit{\textbf{\begin{tabular}[c]{@{}c@{}}Effect\\  Size\end{tabular}}}} \\ \hline
ActiveMQ                           & 6.10                               & 3.47                              & \textless{}0.001     & 0.61                         & 2,632                              & 1,482                              & \textless{}0.001          & 0.19                    & 2.93                              & 1.96                              & \textless{}0.001              & 0.68                \\
Ambari                             & 4.87                              & 3.38                              & \textless{}0.001      & 0.57                        & 2,581                              & 1,683                              & \textless{}0.001           & 0.12                   & 2.87                              & 2.27                              & \textless{}0.001               & 0.56               \\
Arrow                              & 5.34                              & 3.12                              & \textless{}0.001      & 0.51                        & 2,696                              & 1,676                              & \textless{}0.001           & 0.08                   & 2.40                               & 1.72                              & \textless{}0.001              & 0..67                \\
Axis2                              & 5.18                              & 2.86                              & \textless{}0.001      & 0.73                        & 2,261                              & 1,069                              & \textless{}0.001           & 0.41                   & 2.75                              & 1.88                              & \textless{}0.001               & 0.62               \\
Camel                              & 4.72                              & 3.26                              & \textless{}0.001      & 0.40                        & 1,858                              & 1,216                              & \textless{}0.001           & 0.15                  & 2.03                              & 1.69                              & \textless{}0.001                & 0.37             \\
CloudStack                         & 8.69                              & 5.80                              & \textless{}0.001      & 0.25                        & 5,401                              & 4,488                              & \textless{}0.001           & 0.05                   & 3.40                               & 2.14                              & \textless{}0.001              & 0.93                \\
Cordova                            & 6.63                              & 4.56                              & \textless{}0.001      & 0.31                        & 4,067                              & 3,027                              & \textless{}0.001           & 0.02                   & 2.83                              & 2.10                               & \textless{}0.001              & 0.48                \\
Drill                              & 5.61                              & 4.49                              & \textless{}0.001      & 0.16                        & 4,716                              & 6,714                              & \textless{}0.001$\star$      & 0.01                       & 2.50                               & 1.97                              & \textless{}0.001         &  0.45                    \\
Flink                              & 8.94                               & 5.06                              & \textless{}0.001     & 0.52                         & 5,881                              & 3,512                              & \textless{}0.001          & 0.09                    & 3.22                              & 2.05                              & \textless{}0.001              & 0.85                \\
Geode                              & 5.63                              & 4.49                              & 0.005                 & 0.16                        & 7,425                              & 8,363                              & \cellcolor[HTML]{C0C0C0}0.336$\star$   & 0.01             & 2.36                              & 1.78                              & \textless{}0.001           & 0.49                   \\
Groovy                             & 5.39                              & 3.24                              & \textless{}0.001      & 0.56                        & 2,241                              & 1,148                              & \textless{}0.001           & 0.36                   & 2.48                              & 1.83                              & \textless{}0.001               & 0.57               \\
Guacamole                          & 5.24                              & 2.85                              & \textless{}0.001      & 0.52                        & 4,101                              & 1,533                              & \textless{}0.001           & 0.27                   & 2.23                              & 1.59                              & \textless{}0.001               & 0.58               \\
Hadoop                             & 15.8                              & 9.43                              & \textless{}0.001      & 0.60                        & 11,643                             & 7,169                              & \textless{}0.001           & 0.30                   & 4.92                              & 3.53                              & \textless{}0.001               & 0.66               \\
HBase                              & 21.22                             & 11.00                                & \textless{}0.001   & 0.74                           & 16,647                             & 7,587                              & \textless{}0.001        & 0.35                      & 5.11                              & 3.78                              & \textless{}0.001            & 0.62                  \\
Hive                               & 10.49                             & 6.75                              & \textless{}0.001      & 0.49                        & 11,904                             & 9,749                              & \textless{}0.001           & 0.03                   & 3.73                              & 2.82                              & \textless{}0.001               & 0.53               \\
Hudi                               & 2.99                              & 2.94                              & 0.005                 & 0.01                        & 2,933                              & 4,487                              & 0.030$\star$                 & 0.07                        & 1.64                              & 1.28                              & \textless{}0.001          & 0.45                    \\
Ignite                             & 4.92                              & 3.50                               & \textless{}0.001     & 0.34                         & 2,979                              & 1,466                              & \textless{}0.001          & 0.23                    & 2.52                              & 2.10                               & \textless{}0.001            & 0.31                  \\
Impala                             & 5.49                              & 2.91                              & \textless{}0.001      & 0.66                        & 3,418                              & 1,888                              & \textless{}0.001           & 0.23                   & 2.76                              & 1.84                              & \textless{}0.001              & 0.71                \\
Jackrabbit-Oak                     & 7.31                              & 4.47                              & \textless{}0.001      & 0.59                        & 2,244                              & 1,864                              & \textless{}0.001           & 0.02                   & 2.79                              & 2.08                              & \textless{}0.001              & 0.68                \\
Kafka                              & 9.34                              & 4.39                              & \textless{}0.001      & 0.78                        & 7,424                              & 3,672                              & \textless{}0.001           & 0.15                   & 3.86                              & 2.32                              & \textless{}0.001              & 0.77                \\
Lucene                             & 11.49                             & 7.40                               & \textless{}0.001     & 0.47                         & 11,498                             & 2,992                              & \textless{}0.001          & 0.37                    & 3.72                              & 2.84                              & \textless{}0.001             & 0.52                 \\
Mesos                              & 5.86                              & 3.14                              & \textless{}0.001      & 0.72                        & 10,100                             & 2,622                              & \textless{}0.001           & 0.23                   & 2.94                              & 1.85                              & \textless{}0.001             & 0.84                 \\
NetBeans                           & 6.39                              & 2.59                              & \textless{}0.001      & 0.31                        & 7,245                              & 6,746                              & \textless{}0.001           & \textless{}0.01                   & 2.66                              & 1.47                              & \textless{}0.001              & 1.10                \\
NiFi                               & 6.88                              & 4.52                              & \textless{}0.001      & 0.40                        & 4,064                              & 2,702                              & \textless{}0.001          & 0.15                    & 2.72                              & 2.07                              & \textless{}0.001              & 0.52                \\
OFBiz                              & 7.67                              & 4.39                              & \textless{}0.001     & 0.60                         & 3,601                              & 1,481                              & \textless{}0.001          & 0.38                    & 2.96                              & 2.19                              & \textless{}0.001              & 0.65                \\
Ozone                              & 5.04                              & 3.24                              & \textless{}0.001      & 0.38                        & 5,289                              & 3,524                              & \textless{}0.001          & 0.17                    & 2.81                              & 2.07                              & \textless{}0.001              & 0.48                \\
Qpid                               & 4.97                              & 2.93                              & \textless{}0.001      & 0.72                        & 2,012                              & 926                               & \textless{}0.001           & 0.33                   & 2.50                              & 1.90                              & \textless{}0.001               & 0.59               \\
Solr                               & 12.76                             & 6.40                               & \textless{}0.001     & 0.79                         & 5,575                              & 2,738                              & \textless{}0.001         & 0.38                     & 4.18                              & 2.82                              & \textless{}0.001             & 0.72                 \\
Spark                              & 5.93                              & 3.41                               & \textless{}0.001     & 0.62                         & 3,573                              & 1,297                              & \textless{}0.001         & 0.14                     & 3.13                              & 2.17                              & \textless{}0.001             & 0.63                 \\
Thrift                             & 7.31                              & 4.58                              & \textless{}0.001      & 0.48                        & 2,848                              & 1,677                              & \textless{}0.001          & 0.23                    & 2.73                              & 2.41                              & \cellcolor[HTML]{C0C0C0}0.055  & 0.22               \\
Traffic Server                     & 7.81                              & 4.72                              & \textless{}0.001      & 0.40                        & 3,897                              & 2,074                              & \textless{}0.001          & 0.08                    & 3.19                              & 2.16                              & \textless{}0.001               & 0.77               \\
Wicket                             & 6.32                              & 3.91                              & \textless{}0.001      & 0.54                        & 1,981                              & 1,176                              & \textless{}0.001          & 0.27                    & 2.76                              & 2.02                              & \textless{}0.001               & 0.62               \\ \hline
\end{tabular}
}
\end{table*}

\subsubsection{Communication Complexity of Bugs}

Table \ref{table:commentcomplexity} shows the results of the Mann-Whitney U tests on the communication complexity of bugs with priority changes and bugs without priority changes. We explore the communication complexity from the number of comments, the length of comments (the unit of length is the number of bytes), and the number of commenters. As shown in Table \ref{table:commentcomplexity}, for NOC, the AveC is significantly larger than the AveN in all the projects; for TLC and NOCR, the AveC of 29 and 31 of 32 projects respectively are significantly greater than the AveN.




\subsection{Influence of Human Factors on Bug Priority Changes (RQ5)}

\subsubsection{Proportion and Distribution of Different Types of Priority Modifiers}

\begin{figure*}
    \centering{\includegraphics[width=6.2in]{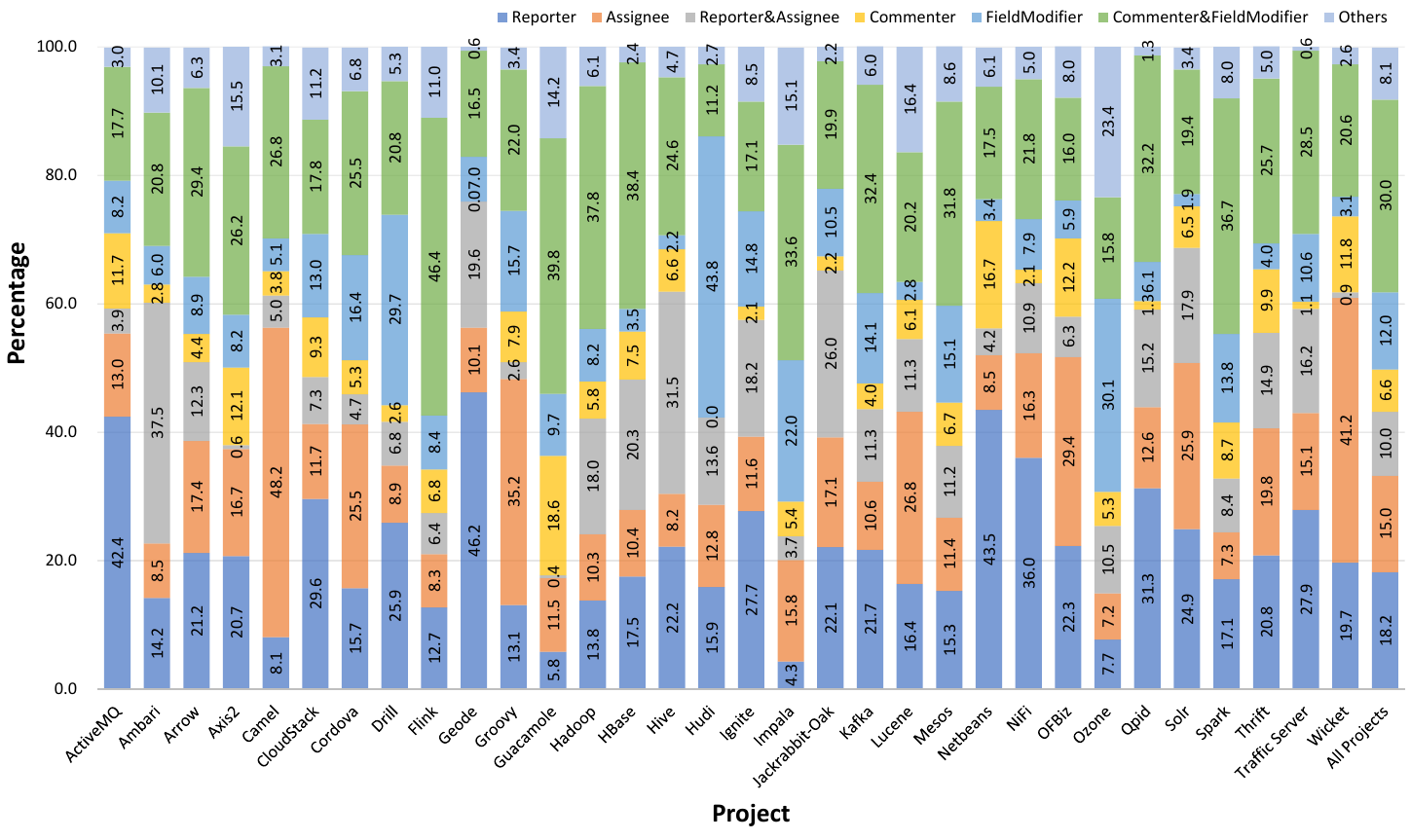}}
    \caption{Distribution of different types of PriorityModifier (RQ5).}
    \label{fig:PersonDis}
\end{figure*}

The proportions of the five types of PriorityModifier are shown in Table \ref{table:persondisdup}, where \# and \% represent the number and percentage, respectively, of PriorityModifier of each type in the corresponding row. 
Taking all projects as a whole, 28.2\% of the PriorityModifiers who have modified the bug priority are Reporter, 25.0\% are Assignee, 71.3\% have commented on the bugs, 81.7\% have modified fields of the bug reports, and only 8.1\% have not performed such activities. Please note that for each row in Table \ref{table:persondisdup}, the sum of the percentages is more than 100\% since every PriorityModifier can belong to multiple types. To accurately describe the distribution of each PriorityModifier type, we further demonstrate the distribution in Figure \ref{fig:PersonDis}, in which the overlaps between different PriorityModifier types are explicitly shown. As we can see from Figure \ref{fig:PersonDis}, the distribution of different types of PriorityModifier in each project varies greatly. For all projects, among the PriorityModifiers who modify the bug priority, 18.2\% are Reporter, 15.0\% are Assignee, 10.0\% are both Reporter and Assignee, 6.6\% have only commented on the corresponding bugs, 12.0\% have only modified fields of the bugs, 30\% have both commented on and modified fields of the bugs, and only 8.1\% have no such activities.

\begin{table*}[]
\centering
\caption{Proportion of different types of PriorityModifier (RQ5).}
\label{table:persondisdup}
\scalebox{0.7}{
\begin{tabular}{lrrrrrr}
\hline
\multirow{2}{*}{\textbf{Project}}      & \multicolumn{1}{c}{\textbf{Reporter}}  & \multicolumn{1}{c}{\textbf{Assignee}}  & \multicolumn{1}{c}{\textbf{Commenter}}  & \multicolumn{1}{c}{\textbf{FieldModifier}} & \multicolumn{1}{c}{\textbf{Others}}   & \multicolumn{1}{c}{\multirow{2}{*}{\textbf{Total}}} \\ \cline{2-6}
                                       & \multicolumn{1}{c}{\# (\%)}                                  & \multicolumn{1}{c}{\# (\%)}                                  & \multicolumn{1}{c}{\# (\%)}                                   & \multicolumn{1}{c}{\# (\%)}                                      & \multicolumn{1}{c}{\# (\%)}                                 & \multicolumn{1}{c}{}                                \\ \hline
ActiveMQ                               & 100 (46.3\%)                             & 39 (16.9\%)                              & 178 (77.1\%)                              & 180 (77.9\%)                                 & 7 ( 3.0\%)                               & 231                                                 \\
Ambari                                 & 164 (51.7\%)                             & 146 (46.1\%)                             & 218 (68.8\%)                              & 270 (85.2\%)                                 & 32 (10.1\%)                             & 317                                                 \\
Arrow                                  & 106 (33.5\%)                             & 94 (29.7\%)                              & 226 (71.5\%)                              & 270 (85.4\%)                                 & 20 ( 6.3\%)                              & 316                                                 \\
Axis2                                  & 107 (21.3\%)                             & 87 (17.3\%)                              & 348 (69.2\%)                              & 330 (65.6\%)                                 & 78 (15.5\%)                             & 503                                                 \\
Camel                                  & 113 (13.0\%)                             & 461 (53.2\%)                             & 738 (85.1\%)                              & 798 (92.0\%)                                 & 27 ( 3.1\%)                              & 867                                                 \\
CloudStack                             & 379 (36.9\%)                             & 195 (19.0\%)                             & 675 (65.7\%)                              & 784 (76.3\%)                                 & 115 (11.2\%)                            & 1,027                                               \\
Cordova                                & 187 (20.4\%)                             & 277 (30.2\%)                             & 629 (68.7\%)                              & 760 (83.0\%)                                 & 62 ( 6.8\%)                              & 916                                                 \\
Drill                                  & 211 (32.8\%)                             & 101 (15.7\%)                             & 355 (55.1\%)                              & 559 (86.8\%)                                 & 34 ( 5.3\%)                              & 644                                                 \\
Flink                                  & 501 (19.1\%)                             & 387 (14.7\%)                             & 2,042 (77.7\%)                            & 2,096 (79.7\%)                               & 288 (11.0\%)                            & 2,629                                               \\
Geode                                  & 104 (65.8\%)                             & 47 (29.7\%)                              & 72 (45.6\%)                               & 144 (91.1\%)                                 & 1 ( 0.6\%)                               & 158                                                 \\
Groovy                                 & 60 (15.7\%)                              & 144 (37.8\%)                             & 275 (72.2\%)                              & 326 (85.6\%)                                 & 13 ( 3.4\%)                              & 381                                                 \\
Guacamole                              & 14 ( 6.2\%)                               & 27 (11.9\%)                              & 162 (71.7\%)                              & 145 (64.2\%)                                 & 32 (14.2\%)                             & 226                                                 \\
Hadoop                                 & 798 (31.8\%)                             & 710 (28.3\%)                             & 2,059 (82.1\%)                            & 2,150 (85.8\%)                               & 152 ( 6.1\%)                             & 2,507                                               \\
HBase                                  & 480 (37.7\%)                             & 390 (30.7\%)                             & 1,168 (91.8\%)                            & 1,129 (88.8\%)                               & 31 ( 2.4\%)                              & 1,272                                               \\
Hive                                   & 295 (53.7\%)                             & 218 (39.7\%)                             & 452 (82.3\%)                              & 462 (84.2\%)                                 & 26 ( 4.7\%)                              & 549                                                 \\
Hudi                                   & 76 (29.5\%)                              & 68 (26.4\%)                              & 75 (29.1\%)                               & 247 (95.7\%)                                 & 7 ( 2.7\%)                               & 258                                                 \\
Ignite                                 & 310 (46.0\%)                             & 201 (29.8\%)                             & 367 (54.5\%)                              & 579 (85.9\%)                                 & 57 ( 8.5\%)                              & 674                                                 \\
Impala                                 & 144 ( 8.0\%)                              & 350 (19.5\%)                             & 1,074 (59.9\%)                            & 1,409 (78.6\%)                               & 270 (15.1\%)                            & 1,792                                               \\
Jackrabbit-Oak                         & 87 (48.1\%)                              & 78 (43.1\%)                              & 148 (81.8\%)                              & 169 (93.4\%)                                 & 4 ( 2.2\%)                               & 181                                                 \\
Kafka                                  & 231 (33.0\%)                             & 153 (21.8\%)                             & 478 (68.2\%)                              & 594 (84.7\%)                                 & 42 ( 6.0\%)                              & 701                                                 \\
Lucene                                 & 59 (27.7\%)                              & 81 (38.0\%)                              & 162 (76.1\%)                              & 163 (76.5\%)                                 & 35 (16.4\%)                             & 213                                                 \\
Mesos                                  & 135 (26.5\%)                             & 115 (22.5\%)                             & 353 (69.2\%)                              & 421 (82.5\%)                                 & 44 ( 8.6\%)                              & 510                                                 \\
Netbeans                               & 180 (47.7\%)                             & 48 (12.7\%)                              & 283 (75.1\%)                              & 232 (61.5\%)                                 & 23 ( 6.1\%)                              & 377                                                 \\
NiFi                                   & 112 (46.9\%)                             & 65 (27.2\%)                              & 152 (63.6\%)                              & 202 (84.5\%)                                 & 12 ( 5.0\%)                              & 239                                                 \\
OFBiz                                  & 68 (28.6\%)                              & 85 (35.7\%)                              & 190 (79.8\%)                              & 183 (76.9\%)                                 & 19 ( 8.0\%)                              & 238                                                 \\
Ozone                                  & 38 (18.2\%)                              & 37 (17.7\%)                              & 77 (36.8\%)                               & 147 (70.3\%)                                 & 49 (23.4\%)                             & 209                                                 \\
Qpid                                   & 107 (46.5\%)                             & 64 (27.8\%)                              & 191 (83.0\%)                              & 214 (93.0\%)                                 & 3 ( 1.3\%)                               & 230                                                 \\
Solr                                   & 177 (42.9\%)                             & 181 (43.8\%)                             & 357 (86.4\%)                              & 355 (86.0\%)                                 & 14 ( 3.4\%)                              & 413                                                 \\
Spark                                  & 517 (25.5\%)                             & 319 (15.7\%)                             & 1,402 (69.1\%)                            & 1,597 (78.7\%)                               & 168 ( 8.0\%)                             & 2,028                                               \\
Thrift                                 & 72 (35.6\%)                              & 70 (34.7\%)                              & 153 (75.7\%)                              & 164 (81.2\%)                                 & 10 ( 5.0\%)                              & 202                                                 \\
Traffic Server                         & 79 (44.1\%)                              & 56 (31.3\%)                              & 131 (73.2\%)                              & 169 (94.4\%)                                 & 1 ( 0.6\%)                               & 179                                                 \\
Wicket                                 & 47 (20.6\%)                              & 96 (42.1\%)                              & 199 (87.3\%)                              & 187 (82.0\%)                                 & 6 ( 2.6\%)                               & 228                                                 \\
\textbf{All Projects} & \textbf{5,267 (28.2\%)} & \textbf{4,680 (25.0\%)} & \textbf{13,330 (71.3\%)} & \textbf{15,285 (81.7\%)}    & \textbf{1,524 ( 8.1\%)} & \textbf{18,708}                    \\ \hline
 \end{tabular}%
 }
 \end{table*}

\subsubsection{Distribution of Bug Participants prone to priority changes}

The numbers of PRs, PAs, and PMs of each project are shown in Table \ref{tab:personprefer}. As we can see, there are 68 PRs in total, scattered in 15 projects, among which \textit{Flink} has the largest number, 27 PRs. There are 13 PAs, which are only distributed in 5 projects, including 7 in \textit{Impala}. There are even less PMs, with only 5 distributed in 5 projects.

\begin{table}[]
\centering
\caption{Distribution of PRs, PAs, and PMs in Selected Projects (RQ5).}
\label{tab:personprefer}
\scalebox{0.7}{
\begin{tabular}{lrrr}
\hline
\textbf{Project}                 & \textbf{\#PR} & \textbf{\#PA} & \textbf{\#PM} \\ \hline
ActiveMQ                         & 0             & 0             & 0             \\
Ambari                           & 0             & 0             & 0             \\
Arrow                            & 0             & 0             & 0             \\
Axis2                            & 2             & 0             & 0             \\
Camel                            & 1             & 0             & 0             \\
CloudStack                       & 5             & 2             & 0             \\
Cordova                          & 1             & 0             & 1             \\
Drill                            & 3             & 0             & 0             \\
Flink                            & 27            & 2             & 1             \\
Geode                            & 0             & 0             & 0             \\
Groovy                           & 0             & 0             & 0             \\
Guacamole                        & 0             & 0             & 1             \\
Hadoop                           & 8             & 0             & 0             \\
HBase                            & 1             & 0             & 0             \\
Hive                             & 0             & 0             & 0             \\
Hudi                             & 3             & 1             & 0             \\
Ignite                           & 0             & 0             & 0             \\
Impala                           & 6             & 7             & 1             \\
Jackrabbit-Oak                   & 0             & 0             & 0             \\
Kafka                            & 2             & 0             & 0             \\
Lucene                           & 0             & 0             & 1             \\
Mesos                            & 2             & 1             & 0             \\
NetBeans                         & 0             & 0             & 0             \\
NiFi                             & 0             & 0             & 0             \\
OFBiz                            & 0             & 0             & 0             \\
Ozone                            & 1             & 0             & 0             \\
Qpid                             & 1             & 0             & 0             \\
Solr                             & 0             & 0             & 0             \\
Spark                            & 5             & 0             & 0             \\
Thrift                           & 0             & 0             & 0             \\
Traffic Server                   & 0             & 0             & 0             \\
Wicket                           & 0             & 0             & 0             \\
\multicolumn{1}{l}{All Projects} & 68            & 13            & 5             \\ \hline
\end{tabular}
}
\end{table}



\section{Discussion}
\label{chap:discus}

In this section, we interpret the study results according to the RQs and discuss the implications of the results for both practitioners and researchers. 
\subsection{Interpretation of Study Results}
\emph{\textbf{RQ1}}: Only a small proportion of bugs experience priority change, as the Apache projects have very clear standards for defining bug priorities \citep{ASF}, and participants are mostly able to correctly evaluate bug priorities based on existing documentation. For some projects, however, the probability of their bugs undergoing priority changes is relatively high compared to the average of all projects, which could be related to the way of bug management. For example, \textit{Flink} has a bot, i.e., Apache Flink Jira Bot, which assists developers in the management of issues. For some other projects, the reason for the high probability of bug priority change may be that the project size is small (e.g., \textit{Guacamole}) or there are batch priority modifications (e.g., \textit{Flink} and \textit{Impala}).

\emph{\textbf{RQ2}}: The minimum time interval for each project is only 8.64 seconds to 345.6 seconds. Due to the fact that the PriorityModifier and bug reporter may be different participants, we only filtered out priority modifications made by the same PriorityModifier within 5 minutes after the bug is reported, while retaining priority modifications made by different PriorityModifiers. The median priority change time interval of bugs is rather short for most projects, while the average value is much longer than the median. This is because in each project there are some bugs with a very long priority change time interval. The relatively small median priority change time interval of the bugs in 28 out of 32 projects indicates that at least a half of the bugs in those projects were reconsidered in terms of bug priority in a few days. The time interval presents this distribution because priority is usually either modified shortly after a bug is reported, or no one claims the bug after it is reported, resulting in the bug to be processed after a long period of time, and the priority changes accordingly.

The priority changes of most bugs in 24 out of 32 projects happened BEFORE. To understand the reasons behind this, we further checked the priority changes that occurred during BEFORE, and the vast majority of the changes occurred in the situation of S4.2 in Figure \ref{fig:changephaseiden} (i.e., the priority changes occurred before the bug was linked to other bugs/issues). Therefore, this to some extent explains the reason for the priority change (i.e., the bug is linked to other bugs), which is consistent with the research results of \cite{AlFeKeSh2020}. 
The reason why the change phase for a small number of bugs is REOPEN is because some bugs start to be fixed after being reopened, and priority changes may also occur. This is a normal phenomenon. However, some bugs still have priority changes after being resolved or closed (i.e., AFTER). Further exploration is needed to uncover the reasons for this phenomenon.


\emph{\textbf{RQ3}}: Among the bugs that have priority changes, most bugs have only one priority change and usually change to adjacent priorities, which resonates with the priority change range of 1 for most bugs. This happens because priority is used to describe the fixing schedule for bugs, and the fixing content represented by the bug itself does not change after it is reported. The fixing work is related to the fixing content, and consequently the times and range of priority changes are relatively small. Accordingly, in the priority change patterns, most bug priority change patterns focus on P1I\_I and P1D\_D. In addition, P2R\_ID and P2R\_DI have the largest number of bugs. This is because among the bugs whose priority is changed twice, the number of bugs that the priority restores finally is the largest. This back and forth modification of the priority is because some PriorityModifier is uncertain about the priority. 

In addition, we are also curious about why some bugs have undergone so many priority changes, and we manually checked the history and comments of bugs whose priorities have changed more than three times. We used open coding and constant comparison of Grounded Theory \citep{StRaFi2016} to generate concepts and categories of reasons for that bugs underwent priority changes more than three times. Grounded Theory (GT) is a bottom-up approach that focuses on theory generation rather than extending or validating existing theory. Open coding and constant comparison are two techniques widely used in Grounded Theory for qualitative data analysis.

The analysis process includes the following steps: (1) The second author used open coding to code the extracted data items (i.e., the comment content of the bug) of bugs with more than three priority changes to generate codes. When the content of the bug comments was unclear and the second author was confused when coding the extracted data, a physical meeting was arranged with the first author to resolve this confusion. (2) The second author applied constant comparison to compare codes identified in one piece of data with codes appearing in other data to identify codes with similar semantics. The second author continued to group similar codes into high-level concepts and categories, and the classification process was iterative, with the second author constantly switching back and forth between categories, content of comments, to modify and refine the categories. (3) Afterwards, the third and fourth authors checked and verified the results of the data analysis (i.e., codes, concepts, and categories). Disagreements were resolved through a meeting using a negotiated agreement approach \citep{CaQuOsPe2013} to improve the reliability of data analysis results. 

In Table \ref{table:reason3}, we summarize the reasons why those bugs underwent priority changes for more than 3 times. To illustrate each reason, we provide a description, an instance, and the number of corresponding bugs. From the distribution of bug numbers corresponding to each reason in Table \ref{table:reason3}, we can see that when a bug undergoes multiple priority changes, the likely reason is that the bug itself is relatively complex, which in turn leads to a more complex process of fixing the bug, increasing uncertainty and affecting the decision of PriorityModifier.

\begin{table*}[]
\centering
\caption{Reasons for bug priority change of more than 3 times (RQ3).}
\label{table:reason3}
\footnotesize
\scalebox{1.0}{
\begin{tabular}{p{0.07\columnwidth} p{0.29\columnwidth} p{0.70\columnwidth} p{0.68\columnwidth} p{0.08\columnwidth}}
\hline
\textbf{\#}   & \textbf{Reason}  & \textbf{Description}  & \textbf{Instance} & \textbf{\#Bug} \\ \hline
\textbf{RS1}    & PriorityModifiers fail to reach agreement with the bug priority. 
& This can be embodied in that two or more modifiers change the priority back and forth in the history, or some commenters raise objections to the priority allocation in the comments.                                                                   & \textit{CLOUDSTACK-1673}, the priority sequence of the bug is 23212. A comment mentioned ``When you mark bug as blocker, please make sure to mention why its blocker''.                                                                                                                                                & 26    \\
\textbf{RS2}    & PriorityModifiers are uncertain about the bug priority. 
& One or more PriorityModifiers change the priority repeatedly, or one or more PriorityModifiers change the priority and then change it back to the original priority.                                                                                    & \textit{IMPALA-1755}, the priority sequence of the bug is 32121. A comment mentioned ``I think it was marked as a blocker since it was a regression, but given that no one seemed to notice I would say downgrade the priority''.                                                                                              & 14    \\
\textbf{RS3}    & The bug priority is version related. 
& For example, a bug is Blocker in the current release but not Blocker in the next release, or it exists in the current release but does not exist in the next release, or the bug priority may be lowered before the next release. & \textit{AXIS2-653}, the priority sequence of the bug is 31312. A comment mentioned ``Reducing the priority as: 1. the discussion on this is yet to be finished; 2. not a blocker for the next release.''                                                                                                                            & 13    \\
\textbf{RS4}    & The bug is unreproducible. 
& After one participant reports the bug, other participants cannot reproduce the bug for some reasons, such as differences in the software environment.                                                                               & \textit{AXIS2-3099}, the priority sequence of the bug is 312124. A comment mentioned that ``Not a blocker as we are not able to reproduce the issue.''                                                                                                                                                                            & 6     \\
\textbf{RS5}    & The bug priority is changed according to the bug fixing progress. 
& The bug is too complex, or the participants do not fully understand the bug, resulting in repeated modification of the priority in the bug-fixing process.                               & \textit{IMPALA-2982}, the priority sequence of the bug is 212134. Two comments mention that ``Bringing this back to Blocker level as the excessive logging is a major supportability issue'' and ``Reduced priority as the estimated impact is not low.'' & 4     \\
\textbf{RS6}    & The bug priority is changed by a Jira bot.  
& There is a bug management robot called Flink Jira Bot used in project \textit{Flink}. It will automatically check the bug progress. If the bug is not updated for a long time, it will decrease the bug priority.                                                                      & \textit{FLINK-18574}, the priority sequence of the bug is 32342. 234 in 32342 is the priority change made by the robot.                                                                                                            & 14    \\
\textbf{RS7}    & The bug priority is changed for unknown reasons.  
& No explicit reasons can be found after checking the history and comments of the bug. This may be because the bug links to other bugs or another instance (including the bugs on JIRA and GitHub, or instance on Azure).                         & \textit{FLINK-17260}, the priority sequence of the bug is 432132. In the comment, the participants associate it with another instance on Azure.                                                                   & 12    \\ \hline
\end{tabular}
}
\end{table*}

For the priority change trend of all bugs, increase and decrease account for the vast majority, with more increase than decrease of priority change. The priority change is also related to the priority itself; the higher the priority, the greater the probability of priority change. The reason for this phenomenon is that participants are usually more cautious about assigning a higher priority to a bug, as a higher priority usually has a greater impact on their work schedule and release planning \citep{ASF}. This leads to the situation that some of such bugs were underestimated with respect to their priority when a bug was reported, and even if a higher priority is assigned, participants are relatively more likely to change their priority.

\emph{\textbf{RQ4}}: Surprisingly, for most projects, the average change complexity of bug-fixing commits for bugs with priority changes is significantly higher than that of bugs without priority changes. This means no matter the trend of priority changes of bugs, the bug-fixing commits for such bugs tend to be more complex than that for bugs without priority changes. In other words, bugs with higher change complexity of bug-fixing commits are more likely to undergo priority changes. We can imagine that it is more difficult to precisely estimate the consequence of a bug with more complex bug-fixing changes. The same is true for the communication complexity. The communication complexity of bugs with priority changes is significantly greater than that of bugs without priority changes. The complexity of comments shows the intensity of the bug discussion among bug participants. In other words, the more comments of bugs, the more likely they are to undergo priority changes. Thus, to some extent, the reason why bugs undergo priority changes is that the fixing of such bugs or bugs themselves is more complex than bugs without priority changes. 

\emph{\textbf{RQ5}}: The study shows that 43.2\% of PriorityModifier are Reporter or Assignee, and most PriorityModifier have commented on the corresponding bug or modified other bug fields except for the Priority field. This is because PriorityModifier is usually familiar with the bug, and also reveals that priority changes may be the result of discussion among PriorityModifier. The proportion of PriorityModifier varies between different projects, as there are significant differences in the size and number of participants of these projects.

Finally, we find that the priority change does have the influence of human factors. However, these numbers are not large, and there are no such participants in many projects, which shows that the impact of human factors is limited. And this seems to be related to the scale of the project. The larger the project, the more participants there will be. The reason for the large number of PRs in project \textit{Flink} is that Flink Jira Bot exists in the project, and there are also some batch modifications, which indicates that the priority change is also related to the development style of the project.

\subsection{Implications for Practitioners}
The results of this study imply a number of points for practitioners, which are presented as follows. 
\begin{itemize}\setlength{\itemsep}{0pt}\setlength{\parskip}{0pt}
\item \textbf{This study helps practitioners have a better understanding of the probability of bug priority change for specific projects.} Due to the varying probability of bug priority change for each project, and even significant differences in some projects, practitioners can evaluate the probability of bug priority change for specific projects based on their characteristics (such as management style and project size), in order to better manage bugs.

\item \textbf{Practitioners need to pay attention to the priority changes that occur at different stages of the bug life cycle.} This study shows that most bug priority changes occur shortly after a bug is reported and before it begins to be processed. Therefore, practitioners can pay more attention to bug priority changes in the early stages of bug fixing to arrange their work schedule reasonably. After the bug is resolved or closed, practitioners should be cautious of changing its priority. Even though the priority change on the resolved or closed bug would not impact the bug fixing process, it may influence the future tasks based on bug priority, such as bug priority prediction and workload estimation.

  
\item \textbf{Bug reports need to display the process of priority changes.} Compared to displaying only the current priority, bug reports showing the process of bug priority changes (such as priority change patterns) can help practitioners understand and fix bugs more efficiently.
\item \textbf{Bugs that require relatively more complex bug-fixing commits or are discussed more during bug fixing tend to undergo priority changes.} This is evidenced by the fact that in most projects, bugs with priority changes have significantly higher code change complexity and communication complexity in bug fixing than bugs without priority changes, as shown in Table \ref{table:changecomplexity} and Table \ref{table:commentcomplexity}, respectively.
\item \textbf{Pay attention to the needs of bug reporters and assignees.} We noticed that a considerable portion of the bug priority changes are modified by the bug reporter and the assignee, indicating that bug reporters and assignees play important roles in the bug fixing process, and understanding the needs of reporters and assignees is crucial for optimizing workflow. Practitioners can understand the needs and expectations of these two roles through regular user feedback and surveys, in order to provide better support and services.
\item \textbf{There is a need to standardize the priority allocation process in order to reduce the impact of human factors during priority allocation.} In some projects, priorities of bugs reported by a few participants or priorities allocated by a small number of participants are more likely to be modified, and a few participants tend to modify priorities. Following a standardized priority allocation process can help to alleviate the impact of potential biases and preferences of such participants in bug priority management.
\end{itemize}

\subsection{Implications for Researchers}
The results of this study also imply a number of points for researchers, which are presented as follows. 
\begin{itemize}\setlength{\itemsep}{0pt}\setlength{\parskip}{0pt}
\item \textbf{This study identifies possible dimensions where researchers can explore the reasons for relatively high priority change probabilities of certain projects.}
Since bugs with priority changes have higher change complexity of bug-fixing commits, it is necessary to make clear the reasons for why bugs in specific projects have high priority change probabilities, in order to find a way to reduce maintenance cost. Our study also identifies several aspects to be further explored by researchers, such as the way of project management (e.g., Flink Jira Bot in \textit{Flink}), project size.

\item \textbf{The phenomenon that a small proportion of bugs occured priority changes after being resloved or closed deserves an in-depth study.} Priority represents the urgency of a bug, and modifying the priority after the bug is resovled or closed does not affect the progress of bug fixing. Therefore, the reasons behind this unreasonable behavior are worth further investigation.
\item \textbf{The bug priority change process may provide useful information to help researchers improve the bug priority prediction models or bug-fixing time prediction models.} Due to the lack of attention paid to bug priority changes, they have hardly been applied to the training of various prediction models. The 24 change patterns proposed in this study characterize the process of priority changes for each bug, which may help to improve the performance of the prediction models.
\item \textbf{Further investigation on the reasons for bug priority changes is needed.} Currently, the reasons have merely been investigated from the perspective of the change complexity of bug-fixing commits and the communication complexity of bugs. Other perspectives, such as project-specific factors, should also be further studied. To estimate bug priorities more accurately thereby optimizing release planning and task assignment in project management, it is worthwhile to look into diverse factors that drive bug priority changes in depth.
\item \textbf{It is worthwhile to conduct in-depth research on the role of the main participants in bug fixing.} Our research indicates that a significant portion of PriorityModifiers are bug reporters and assignees, and researchers can gain a deeper understanding of their specific roles in bug discovery, resolution, and validation to better understand the dynamics of bug fixing.
\end{itemize}

\section{Threats to Validity}
\label{chap:threats}

There are several threats to the validity of the study results. We discuss these threats according to the guidelines in \citep{RuHo2009}. Please note that internal validity is not discussed since we did not study causal relationships.

\subsection{Construct Validity}
Construct validity is concerned with whether the values of the variables (listed in TABLE \ref{table:dataitem}) we obtained are in line with the real values that we expected. A potential threat to construct validity is that not all bugs resolved are linked to the corresponding commits. Due to different developer habits and development cultures, committers may not explicitly mention the ID of the bug resolved in the corresponding commit message, which may negatively affect the representativeness of the collected bugs and further influence the accuracy of defect density and the time taken to resolve bugs. Through our analysis (the analysis results are not shown in this paper due to its deviation from the focus of this paper), we confirmed that the committers who explicitly mention the bug ID do not come from a small group of specific developers. Therefore, this threat is mitigated to some extent. 


Another threat is that a bug will probably be reopened and the priority may be changed again, which results in the incompleteness of priority changes of the bug and thus affects the study results. We analyzed the data of the bugs that were resolved or closed before December 1, 2019 and were not reopened again before February 4, 2021. It means that the status of such bugs had been stable for one year and two months after they were resolved or closed. We believe that the likelihood of such bugs being reopened is significantly decreased. Therefore, this threat is alleviated to a large extent.

\subsection{External Validity}
External validity is concerned with the generalizability of the study results.
First, we only consider the five bug priorities of Blocker, Critical, Major, Minor, and Trivial in JIRA. The research results may not be generalized to other priorities in JIRA or priorities in other issue tracking systems.
Then another potential threat to external validity is whether the selected projects are representative enough. As presented in Section \ref{CaseSelection}, we applied a set of criteria to select projects. We tried to include as many Apache projects that meet the selection criteria as possible. 
Furthermore, the selected projects cover different languages and different application domains, and differ in code repository size and development duration. This indicates an improved representativeness of the selected projects.
Finally, since only OSS projects were selected, the study results and findings may not be generalized to closed source software projects.

\subsection{Reliability}
Reliability refers to whether the study yields the same results when replicated by other researchers. A potential threat is related to the implementation of related software tools for data collection. The tools were mainly implemented by the third author, and the code of the key functionalities had been regularly reviewed by the first and second authors. Furthermore, sufficient tests were performed to ensure the correctness of the calculation of data items. Hence, this threat to reliability had been alleviated.

Another threat is related to the correctness of the Mann-Whitney U tests. Since we used IBM SPSS (a widely-used professional tool for statistics) and the Scipy library in Python (a scientific computing-related package of Python) to run the tests. Hence, we believe that the threat is minimized.

\section{Conclusions and Future Work}\label{conclusions}
In this work, we investigated the phenomenon of bug priority changes by analyzing bugs and related data collected from 32 non-trivial OSS projects of Apache Software Foundation. We identified 24 patterns characterizing the process of bug priority changes. The main findings are summarized as follows:
 \begin{itemize}\setlength{\itemsep}{0pt}\setlength{\parskip}{0pt}
  \item A small proportion (only 8.3\%) of bugs of the selected projects undergo priority changes.
  \item In 28 out of 32 projects, the median time interval of priority changes is less than ten days.
  \item In 24 out of 32 projects, the number of bugs whose priority changes happened before the bug-fixing process accounts for the largest. 
  \item At least 81.6\% of the bugs with priority changes of each project undergo only one priority change, and no more than 8.2\% of the bugs with priority changes of each project undergo three or more priority changes.
  \item Each project covers 6 to 22 bug priority change patterns, none of the projects covers all the 24 patterns, and most of the bugs with priority changes increase the bug priority finally.
  \item Most priority changes tend to shift the priority to its adjacent priority and a higher priority holds a greater probability to undergo priority change.
  \item In most of the projects, bugs that undergo priority changes have significantly higher change complexity of bug-fixing commits and communication complexity than bugs that do not undergo priority changes.
  \item 43.2\% of the participants who make priority changes are the reporter or assignee of the bug, 71.3\% of them make comments on the bug, and 81.7\% of them modify other fields of the bug except the priority field.
  \item In around a half of the 32 projects, bugs reported by a few participants are more likely to be modified; in 5 out of the 32 projects, the priorities allocated by a few participants are more likely to be modified; and in 5 out of the 32 projects, there is one participant who is prone to modifying the bug priority.
\end{itemize}

Based on the results of this study, our future research will focus on the following directions: 
first, to investigate bug priority changes in other OSS ecosystems using different issue tracking systems with distinct priority models; 
second, to study the relationship between priority changes and different issue types or project types; 
third, to explore how human factors specifically affect bug priority changes.

\section*{Data availability}
We have shared the link to our dataset in the reference \citep{dataset}.


\section*{Acknowledgments}
This work is supported by the Natural Science Foundation of Hubei Province of China under Grant No. 2021CFB577, the National Natural Science Foundation of China under Grant Nos. 62176099 and 62172311, and the Knowledge Innovation Program of Wuhan-Shuguang Project under Grant No. 2022010801020280.

\printcredits


\bibliographystyle{cas-model2-names}

\bibliography{references}

\begin{thebibliography}{35}
\expandafter\ifx\csname natexlab\endcsname\relax\def\natexlab#1{#1}\fi
\providecommand{\url}[1]{\texttt{#1}}
\providecommand{\href}[2]{#2}
\providecommand{\path}[1]{#1}
\providecommand{\DOIprefix}{doi:}
\providecommand{\ArXivprefix}{arXiv:}
\providecommand{\URLprefix}{URL: }
\providecommand{\Pubmedprefix}{pmid:}
\providecommand{\doi}[1]{\href{http://dx.doi.org/#1}{\path{#1}}}
\providecommand{\Pubmed}[1]{\href{pmid:#1}{\path{#1}}}
\providecommand{\bibinfo}[2]{#2}
\ifx\xfnm\relax \def\xfnm[#1]{\unskip,\space#1}\fi
\bibitem[{Akbarinasaji et~al.(2018)Akbarinasaji, Caglayan and Bener}]{AkCaBe2018}
\bibinfo{author}{Akbarinasaji, S.}, \bibinfo{author}{Caglayan, B.}, \bibinfo{author}{Bener, A.}, \bibinfo{year}{2018}.
\newblock \bibinfo{title}{Predicting bug-fixing time: A replication study using an open source software project}.
\newblock \bibinfo{journal}{Journal of Systems and Software} \bibinfo{volume}{136}, \bibinfo{pages}{173--186}.
\bibitem[{Al-Sabbagh et~al.(2022)Al-Sabbagh, Staron and Hebig}]{AlStHe2022}
\bibinfo{author}{Al-Sabbagh, K.}, \bibinfo{author}{Staron, M.}, \bibinfo{author}{Hebig, R.}, \bibinfo{year}{2022}.
\newblock \bibinfo{title}{Predicting build outcomes in continuous integration using textual analysis of source code commits}, in: \bibinfo{booktitle}{Proceedings of the 18th International Conference on Predictive Models and Data Analytics in Software Engineering}, pp. \bibinfo{pages}{42--51}.
\bibitem[{Alenezi and Banitaan(2013)}]{AlBa2014}
\bibinfo{author}{Alenezi, M.}, \bibinfo{author}{Banitaan, S.}, \bibinfo{year}{2013}.
\newblock \bibinfo{title}{Bug reports prioritization: Which features and classifier to use?}, in: \bibinfo{booktitle}{2013 12th International Conference on Machine Learning and Applications}, pp. \bibinfo{pages}{112--116}.
\bibitem[{{Almhana} et~al.(2020){Almhana}, {Ferreira}, {Kessentini} and {Sharma}}]{AlFeKeSh2020}
\bibinfo{author}{{Almhana}, R.}, \bibinfo{author}{{Ferreira}, T.}, \bibinfo{author}{{Kessentini}, M.}, \bibinfo{author}{{Sharma}, T.}, \bibinfo{year}{2020}.
\newblock \bibinfo{title}{Understanding and characterizing changes in bugs priority: The practitioners’ perceptive}, in: \bibinfo{booktitle}{Proceedings of the 20th International Working Conference on Source Code Analysis and Manipulation (SCAM'20)}, \bibinfo{publisher}{IEEE}. pp. \bibinfo{pages}{87--97}.
\bibitem[{Apache(2023)}]{ASF}
\bibinfo{author}{Apache}, \bibinfo{year}{2023}.
\newblock \bibinfo{title}{“guidelines for creating a jira ticket”}.
\newblock \URLprefix \url{https://infra.apache.org/jira-guidelines.html}. \bibinfo{note}{accessed: January 22, 2023}.
\bibitem[{Basili(1992)}]{Ba1992}
\bibinfo{author}{Basili, V.R.}, \bibinfo{year}{1992}.
\newblock \bibinfo{title}{Software modeling and measurement: The goal/question/metric paradigm}.
\newblock \URLprefix \url{http://drum.lib.umd.edu/bitstream/1903/7538/1/Goal_Question_Metric.pdf}.
\bibitem[{Campbell et~al.(2013)Campbell, Quincy, Osserman and Pedersen}]{CaQuOsPe2013}
\bibinfo{author}{Campbell, J.L.}, \bibinfo{author}{Quincy, C.}, \bibinfo{author}{Osserman, J.}, \bibinfo{author}{Pedersen, O.K.}, \bibinfo{year}{2013}.
\newblock \bibinfo{title}{Coding in-depth semistructured interviews: Problems of unitization and intercoder reliability and agreement}.
\newblock \bibinfo{journal}{Sociological methods \& research} \bibinfo{volume}{42}, \bibinfo{pages}{294--320}.
\bibitem[{Chauhan and Kumar(2020)}]{ChKu2020}
\bibinfo{author}{Chauhan, A.}, \bibinfo{author}{Kumar, R.}, \bibinfo{year}{2020}.
\newblock \bibinfo{title}{Bug severity classification using semantic feature with convolution neural network}, in: \bibinfo{booktitle}{Computing in Engineering and Technology}. \bibinfo{publisher}{Springer}, pp. \bibinfo{pages}{327--335}.
\bibitem[{Cheng et~al.(2017)Cheng, Li, Li, Zhao and Liao}]{ChLiLiZhLi2017}
\bibinfo{author}{Cheng, C.}, \bibinfo{author}{Li, B.}, \bibinfo{author}{Li, Z.Y.}, \bibinfo{author}{Zhao, Y.Q.}, \bibinfo{author}{Liao, F.L.}, \bibinfo{year}{2017}.
\newblock \bibinfo{title}{Developer role evolution in open source software ecosystem: An explanatory study on gnome}.
\newblock \bibinfo{journal}{Journal of computer science and technology} \bibinfo{volume}{32}, \bibinfo{pages}{396--414}.
\bibitem[{Etemadi et~al.(2021)Etemadi, Bushehrian, Akbari and Robles}]{EtBuAkRo2021}
\bibinfo{author}{Etemadi, V.}, \bibinfo{author}{Bushehrian, O.}, \bibinfo{author}{Akbari, R.}, \bibinfo{author}{Robles, G.}, \bibinfo{year}{2021}.
\newblock \bibinfo{title}{A scheduling-driven approach to efficiently assign bug fixing tasks to developers}.
\newblock \bibinfo{journal}{Journal of Systems and Software} \bibinfo{volume}{178}, \bibinfo{pages}{110967}.
\bibitem[{Feng et~al.(2012)Feng, Khomh, Ying and Hassan}]{FeKhYiHa2012}
\bibinfo{author}{Feng, Z.}, \bibinfo{author}{Khomh, F.}, \bibinfo{author}{Ying, Z.}, \bibinfo{author}{Hassan, A.E.}, \bibinfo{year}{2012}.
\newblock \bibinfo{title}{An empirical study on factors impacting bug fixing time}, in: \bibinfo{booktitle}{Reverse Engineering}.
\bibitem[{Field(2013)}]{Fi2013}
\bibinfo{author}{Field, A.}, \bibinfo{year}{2013}.
\newblock \bibinfo{title}{Discovering Statistics using IBM SPSS Statistics}.
\newblock \bibinfo{edition}{Fourth} ed., \bibinfo{publisher}{Sage Publications Ltd.}, \bibinfo{address}{Singapore}.
\bibitem[{Gavidia-Calderon et~al.(2021)Gavidia-Calderon, Sarro, Harman and Barr}]{GaSaHaBa2019}
\bibinfo{author}{Gavidia-Calderon, C.}, \bibinfo{author}{Sarro, F.}, \bibinfo{author}{Harman, M.}, \bibinfo{author}{Barr, E.T.}, \bibinfo{year}{2021}.
\newblock \bibinfo{title}{The assessor's dilemma: Improving bug repair via empirical game theory}.
\newblock \bibinfo{journal}{IEEE Transactions on Software Engineering} \bibinfo{volume}{47}, \bibinfo{pages}{2143--2161}.
\bibitem[{G{\"o}k{\c{c}}eo{\u{g}}lu and S{\"o}zer(2021)}]{GoSo2021}
\bibinfo{author}{G{\"o}k{\c{c}}eo{\u{g}}lu, M.}, \bibinfo{author}{S{\"o}zer, H.}, \bibinfo{year}{2021}.
\newblock \bibinfo{title}{Automated defect prioritization based on defects resolved at various project periods}.
\newblock \bibinfo{journal}{Journal of Systems and Software} \bibinfo{volume}{179}, \bibinfo{pages}{110993}.
\bibitem[{Habayeb et~al.(2018)Habayeb, Murtaza, Miranskyy and Bener}]{HaMuMiBe2017}
\bibinfo{author}{Habayeb, M.}, \bibinfo{author}{Murtaza, S.S.}, \bibinfo{author}{Miranskyy, A.}, \bibinfo{author}{Bener, A.B.}, \bibinfo{year}{2018}.
\newblock \bibinfo{title}{On the use of hidden markov model to predict the time to fix bugs}.
\newblock \bibinfo{journal}{IEEE Transactions on Software Engineering} \bibinfo{volume}{44}, \bibinfo{pages}{1224--1244}.
\bibitem[{Hassan(2009)}]{Ha2009}
\bibinfo{author}{Hassan, A.E.}, \bibinfo{year}{2009}.
\newblock \bibinfo{title}{Predicting faults using the complexity of code changes}, in: \bibinfo{booktitle}{Proceedings of the 31st International Conference on Software Engineering (ICSE'09)}, \bibinfo{publisher}{IEEE}. pp. \bibinfo{pages}{78--88}.
\bibitem[{Izadi et~al.(2022)Izadi, Akbari and Heydarnoori}]{IzAkHe2022}
\bibinfo{author}{Izadi, M.}, \bibinfo{author}{Akbari, K.}, \bibinfo{author}{Heydarnoori, A.}, \bibinfo{year}{2022}.
\newblock \bibinfo{title}{Predicting the objective and priority of issue reports in software repositories}.
\newblock \bibinfo{journal}{Empirical Software Engineering} \bibinfo{volume}{27}, \bibinfo{pages}{50}.
\bibitem[{Kanwal and Maqbool(2012)}]{KaMa2012}
\bibinfo{author}{Kanwal, J.}, \bibinfo{author}{Maqbool, O.}, \bibinfo{year}{2012}.
\newblock \bibinfo{title}{Bug prioritization to facilitate bug report triage}.
\newblock \bibinfo{journal}{Journal of Computer Science and Technology} \bibinfo{volume}{27}, \bibinfo{pages}{397--412}.
\bibitem[{Kononenko et~al.(2016)Kononenko, Baysal and Godfrey}]{KoBaGo2016}
\bibinfo{author}{Kononenko, O.}, \bibinfo{author}{Baysal, O.}, \bibinfo{author}{Godfrey, M.W.}, \bibinfo{year}{2016}.
\newblock \bibinfo{title}{Code review quality: How developers see it}, in: \bibinfo{booktitle}{Proceedings of the 38th international conference on software engineering}, pp. \bibinfo{pages}{1028--1038}.
\bibitem[{Kumari and Singh(2020)}]{KuSi2020}
\bibinfo{author}{Kumari, M.}, \bibinfo{author}{Singh, V.B.}, \bibinfo{year}{2020}.
\newblock \bibinfo{title}{An improved classifier based on entropy and deep learning for bug priority prediction}, in: \bibinfo{booktitle}{Abraham, A., Cherukuri, A., Melin, P., Gandhi, N. (eds) Intelligent Systems Design and Applications (ISDA'18)}, pp. \bibinfo{pages}{571--580}.
\bibitem[{Li et~al.(2024)Li, Cai, Yu, Liang, Mo and Liu}]{dataset}
\bibinfo{author}{Li, Z.}, \bibinfo{author}{Cai, G.}, \bibinfo{author}{Yu, Q.}, \bibinfo{author}{Liang, P.}, \bibinfo{author}{Mo, R.}, \bibinfo{author}{Liu, H.}, \bibinfo{year}{2024}.
\newblock \bibinfo{title}{Replication package for “bug priority change: An empirical study on apache projects”}.
\newblock \URLprefix \url{https://github.com/breezesway/BugPriorityChange}.
\bibitem[{Li et~al.(2020)Li, Liang, Li, Mo and Li}]{LiLiLiMoLi2020}
\bibinfo{author}{Li, Z.}, \bibinfo{author}{Liang, P.}, \bibinfo{author}{Li, D.}, \bibinfo{author}{Mo, R.}, \bibinfo{author}{Li, B.}, \bibinfo{year}{2020}.
\newblock \bibinfo{title}{Is bug severity in line with bug fixing change complexity?}
\newblock \bibinfo{journal}{International Journal of Software Engineering and Knowledge Engineering} \bibinfo{volume}{30}, \bibinfo{pages}{1779--1800}.
\bibitem[{Menzies and Marcus(2008)}]{MeMa2008}
\bibinfo{author}{Menzies, T.}, \bibinfo{author}{Marcus, A.}, \bibinfo{year}{2008}.
\newblock \bibinfo{title}{Automated severity assessment of software defect reports}, in: \bibinfo{booktitle}{2008 IEEE International Conference on Software Maintenance}, pp. \bibinfo{pages}{346--355}.
\bibitem[{Motwani et~al.(2018)Motwani, Sankaranarayanan, Just and Brun}]{MoSaJuBr2018}
\bibinfo{author}{Motwani, M.}, \bibinfo{author}{Sankaranarayanan, S.}, \bibinfo{author}{Just, R.}, \bibinfo{author}{Brun, Y.}, \bibinfo{year}{2018}.
\newblock \bibinfo{title}{Do automated program repair techniques repair hard and important bugs?}
\newblock \bibinfo{journal}{Empirical Software Engineering} \bibinfo{volume}{23}, \bibinfo{pages}{2901--2947}.
\bibitem[{Najafi et~al.(2019)Najafi, Rigby and Shang}]{NaRiSh2019}
\bibinfo{author}{Najafi, A.}, \bibinfo{author}{Rigby, P.C.}, \bibinfo{author}{Shang, W.}, \bibinfo{year}{2019}.
\newblock \bibinfo{title}{Bisecting commits and modeling commit risk during testing}, in: \bibinfo{booktitle}{Proceedings of the 2019 27th ACM Joint Meeting on European Software Engineering Conference and Symposium on the Foundations of Software Engineering}, pp. \bibinfo{pages}{279--289}.
\bibitem[{Oliveira et~al.(2020)Oliveira, Fernandes, Steinmacher, Cristo, Conte and Garcia}]{OlFeStCrCoGa2020}
\bibinfo{author}{Oliveira, E.}, \bibinfo{author}{Fernandes, E.}, \bibinfo{author}{Steinmacher, I.}, \bibinfo{author}{Cristo, M.}, \bibinfo{author}{Conte, T.}, \bibinfo{author}{Garcia, A.}, \bibinfo{year}{2020}.
\newblock \bibinfo{title}{Code and commit metrics of developer productivity: a study on team leaders perceptions}.
\newblock \bibinfo{journal}{Empirical Software Engineering} \bibinfo{volume}{25}, \bibinfo{pages}{2519--2549}.
\bibitem[{Runeson and H{\"o}st(2009)}]{RuHo2009}
\bibinfo{author}{Runeson, P.}, \bibinfo{author}{H{\"o}st, M.}, \bibinfo{year}{2009}.
\newblock \bibinfo{title}{Guidelines for conducting and reporting case study research in software engineering}.
\newblock \bibinfo{journal}{Empirical Software Engineering} \bibinfo{volume}{14}, \bibinfo{pages}{131--164}.
\bibitem[{Sharma et~al.(2012)Sharma, Bedi, Chaturvedi and Singh}]{ShBeChSi2013}
\bibinfo{author}{Sharma, M.}, \bibinfo{author}{Bedi, P.}, \bibinfo{author}{Chaturvedi, K.}, \bibinfo{author}{Singh, V.}, \bibinfo{year}{2012}.
\newblock \bibinfo{title}{Predicting the priority of a reported bug using machine learning techniques and cross project validation}, in: \bibinfo{booktitle}{2012 12th International Conference on Intelligent Systems Design and Applications (ISDA)}, pp. \bibinfo{pages}{539--545}.
\bibitem[{Stol et~al.(2016)Stol, Ralph and Fitzgerald}]{StRaFi2016}
\bibinfo{author}{Stol, K.J.}, \bibinfo{author}{Ralph, P.}, \bibinfo{author}{Fitzgerald, B.}, \bibinfo{year}{2016}.
\newblock \bibinfo{title}{Grounded theory in software engineering research: a critical review and guidelines}, in: \bibinfo{booktitle}{Proceedings of the 38th International conference on software engineering}, pp. \bibinfo{pages}{120--131}.
\bibitem[{Tian et~al.(2016)Tian, Ali, Lo and Hassan}]{TiAlLo2016}
\bibinfo{author}{Tian, Y.}, \bibinfo{author}{Ali, N.}, \bibinfo{author}{Lo, D.}, \bibinfo{author}{Hassan, A.E.}, \bibinfo{year}{2016}.
\newblock \bibinfo{title}{On the unreliability of bug severity data}.
\newblock \bibinfo{journal}{Empirical Software Engineering} \bibinfo{volume}{21}, \bibinfo{pages}{2298–2323}.
\bibitem[{Vieira et~al.(2022)Vieira, Mattos, Rocha, Gomes and Paix{\~a}o}]{ViMaRoGoPa2022}
\bibinfo{author}{Vieira, R.G.}, \bibinfo{author}{Mattos, C.L.C.}, \bibinfo{author}{Rocha, L.S.}, \bibinfo{author}{Gomes, J.P.P.}, \bibinfo{author}{Paix{\~a}o, M.}, \bibinfo{year}{2022}.
\newblock \bibinfo{title}{The role of bug report evolution in reliable fixing estimation}.
\newblock \bibinfo{journal}{Empirical Software Engineering} \bibinfo{volume}{27}, \bibinfo{pages}{1--39}.
\bibitem[{Yu et~al.(2010)Yu, Tsai, Zhao and Wu}]{YuTsZhWu2010}
\bibinfo{author}{Yu, L.}, \bibinfo{author}{Tsai, W.T.}, \bibinfo{author}{Zhao, W.}, \bibinfo{author}{Wu, F.}, \bibinfo{year}{2010}.
\newblock \bibinfo{title}{Predicting defect priority based on neural networks}, in: \bibinfo{booktitle}{International Conference on Advanced Data Mining and Applications}, \bibinfo{organization}{Springer}. pp. \bibinfo{pages}{356--367}.
\bibitem[{Yuan et~al.(2015)Yuan, Lo, Xin and Sun}]{YuLoXiSu2015}
\bibinfo{author}{Yuan, T.}, \bibinfo{author}{Lo, D.}, \bibinfo{author}{Xin, X.}, \bibinfo{author}{Sun, C.}, \bibinfo{year}{2015}.
\newblock \bibinfo{title}{Automated prediction of bug report priority using multi-factor analysis}.
\newblock \bibinfo{journal}{Empirical Software Engineering} \bibinfo{volume}{20}, \bibinfo{pages}{1354--1383}.
\bibitem[{Yuan et~al.(2020)Yuan, Lu, Sun and Liu}]{YuLuSuLi2020}
\bibinfo{author}{Yuan, W.}, \bibinfo{author}{Lu, S.}, \bibinfo{author}{Sun, H.}, \bibinfo{author}{Liu, X.}, \bibinfo{year}{2020}.
\newblock \bibinfo{title}{How are distributed bugs diagnosed and fixed through system logs?}
\newblock \bibinfo{journal}{Information and Software Technology} \bibinfo{volume}{119}, \bibinfo{pages}{106234}.
\bibitem[{Zou et~al.(2018)Zou, Lo, Chen, Xia, Feng and Xu}]{ZoLoChXiFeXu2018}
\bibinfo{author}{Zou, W.}, \bibinfo{author}{Lo, D.}, \bibinfo{author}{Chen, Z.}, \bibinfo{author}{Xia, X.}, \bibinfo{author}{Feng, Y.}, \bibinfo{author}{Xu, B.}, \bibinfo{year}{2018}.
\newblock \bibinfo{title}{How practitioners perceive automated bug report management techniques}.
\newblock \bibinfo{journal}{IEEE Transactions on Software Engineering} \bibinfo{volume}{46}, \bibinfo{pages}{836--862}.

\end{thebibliography}
\balance
\end{sloppypar}
\end{document}